\documentclass[12pt]{article}

\usepackage{amsfonts,amssymb,amsmath,enumerate}
\usepackage{color}
\usepackage{theorem,cite}
\usepackage{indentfirst}
\usepackage{textcomp}

\newtheorem{prop}{Proposition}[section]

\theorembodyfont{\rmfamily}
\newtheorem{rmk}{Remark}[section]

\newcommand{\bea}{\begin{eqnarray}}
\newcommand{\eea}{\end{eqnarray}}
\newcommand{\shit}{\begin{align}}
\newcommand{\mdr}{\end{align}}
\newcommand{\beano}{\begin{eqnarray*}}
\newcommand{\eeano}{\end{eqnarray*}}
\newcommand{\beq}{\begin{equation}}
\newcommand{\eeq}{\end{equation}}

\newcommand{\mb}[1]{\hspace{2.1ex}\mbox{#1}\hspace{2.1ex}}
\numberwithin{equation}{section}

\newcommand{\eps}{\epsilon}

\newcommand{\vph}{\varphi}

    \def\cE{{\cal E}}    
    \def\cH{{\cal H}}    
        
    \def\cN{{\cal N}}    
        
        \def\cU{{\cal U}}
\def\cV{{\cal V}}

\newcommand{\CC}{{\mathbb C}}

\newcommand{\II}{{\mathbb I}}

\newcommand{\wt}[1]{\widetilde{#1}}

\newcommand{\half}{{\textstyle{\frac{1}{2}}}}

\def\diag{\mathop{\rm diag}\nolimits}

\begin{document}
%%%%%%%%%%%%%%%%%%%%%%%%%%%%%%%%%%%%%%%%%%%%%%%%%%%%%%%%%%%%%%%
\pagestyle{empty}
\begin{flushright}
LAPTH-035/13
\end{flushright}

\vspace{20pt}

\begin{center}
\begin{LARGE}
{\bf Classification of three-state Hamiltonians\\[1.2ex] solvable by Coordinate Bethe Ansatz}
\end{LARGE}

\vspace{50pt}

\begin{large}
{N.~Cramp\'e${}^{a,b}$, L.~Frappat${}^c$, E.~Ragoucy${}^c$ \footnote[1]{nicolas.crampe@univ-montp2.fr, luc.frappat@lapth.cnrs.fr, eric.ragoucy@lapth.cnrs.fr}}
\end{large}

\vspace{15mm}

${}^a$ {\it CNRS, Laboratoire Charles Coulomb L2C  UMR 5221,\\
Place Eug\`ene Bataillon - CC069, F-34095 Montpellier Cedex 5, France}

\vspace{5mm}

${}^b$ {\it Universit\'e Montpellier II, Laboratoire Charles Coulomb,\\
Place Eug\`ene Bataillon - CC069, F-34095 Montpellier Cedex 5, France}

\vspace{5mm}

${}^c$ {\it Laboratoire de Physique Th\'eorique LAPTH, CNRS and Universit\'e de Savoie,\\
BP 110, 74941 Annecy-le-Vieux Cedex, France}

\end{center}

\vspace{4mm}

\begin{abstract}
We classify `all' Hamiltonians with rank 1 symmetry and nearest neighbour interactions, acting on a periodic three-state spin chain, and solvable through (generalization of) the coordinate Bethe ansatz (CBA). We obtain in this way four multi-parametric extensions of the known 19-vertex Hamiltonians (such as Zamolodchikov-Fateev, Izergin-Korepin, Bariev Hamiltonians). 
Apart from the 19-vertex Hamiltonians, there exists 17-vertex and 14-vertex Hamiltonians that cannot be viewed as subcases of the 19-vertex ones. In the case of 17-vertex Hamiltonian, we get a generalization of the genus 5 special branch found by Martins, plus three new ones. We get also two 14-vertex Hamiltonians.

We solve all these Hamiltonians using CBA, and provide their spectrum, eigenfunctions and Bethe equations. A special attention is made to provide the specifications of our multi-parametric Hamiltonians that give back known Hamiltonians.
\end{abstract}

%%%%%%%%%%%%%%%%%%%%%%%%%%%%%%%%%%%%%%%%%%%%%%%%%%%%%%%%%%%%%%%%%

\newpage
\pagestyle{plain}
\setcounter{page}{1}

\section{Introduction}

In his seminal paper \cite{Bethe}, Bethe provided a method, called now coordinate Bethe ansatz (CBA), 
to compute the eigenvalues and the eigenfunctions for the Heisenberg (or XXX 1/2-spin) models \cite{Heis}.
The same idea has been also used intensively in the context of the Bose gas with $\delta$-interaction 
when the particles have no spin and with periodic boundary condition \cite{LL}. When they carry a spin  
\cite{Yang,Suth}, the famous Yang-Baxter equation shows up through the same techniques, and actually
it appeared for the first time in this context.  
When open boundaries are imposed, the procedure needs to be modified but still applies \cite{Gaudin,bach}, 
even when the boundaries are not diagonal \cite{nondiag}. The Hubbard model \cite{Hub} is another example 
where the CBA has been used successfully \cite{LW}. Remark also that, depending on the context,  different generalizations of CBA have been 
considered, see e.g. \cite{LS, CBA-TL, CRA, TL-open}.

However this method has been considered as deprecated in the eighties in favor of
the quantum inverse scattering method (QISM) \cite{kusk,tafa}.
This latter method is more algebraic and provides a full set of commuting Hamiltonians. It lies on the 
explicit numerical solution of the Yang-Baxter equation and on  representations  of the underlying algebra.
Nevertheless, to the best of our knowledge, one does not know, in general, if the set of models solved by QISM 
and the ones solved by CBA are equivalent.     

In this paper, we classify the most general Hamiltonians with nearest neighbourg interaction and acting on a three-state spin chain 
with rank 1 symmetry (i.e. the nineteen-vertex models) that can be solved by the CBA. We also compare 
our results with the classification of the solutions of the Yang-Baxter equation for the
nineteen vertex model \cite{ITA,MP,M13}. To solve this type of models by CBA, the historical method 
must be generalized following the lines of \cite{BNW,LS} where particular 19-vertex models\footnote{See remark \ref{rmk:19v} for the name ``19-vertex''.}
(the Izergin-Korepin and the Zamolodchikov-Fateev models) or of \cite{CRA} where higher spin chain have 
been solved. This type of computation has been initiated in \cite{AB} but the huge algebraic 
equations the authors got did not allow them to provide a classification. Here, with the use 
of formal mathematical software, we succeed in obtaining a complete classification.
We recover as subcases all the models discovered by solving the Yang-Baxter equation and, as an important by-product, we get the eigenvalues
and eigenfunctions, which were not known previously, for the models obtained in \cite{M13} (see section \ref{sec:19V}).
We also obtain 
four 17-vertex models, one of them being a generalization of the special branch found in \cite{M13}, and two new 14-vertex models. These 17-vertex and 14-vertex Hamiltonians, are \textit{not} subcases of 19-vertex Hamiltonians.

The paper is organized as follows. In section \ref{sec:GS}, we present the general Hamiltonian we want to 
solve and the symmetries one may consider. We give the outlines of the coordinate Bethe ansatz in 
section \ref{sec:CBA}: we derive the Bethe ansatz equations (BAE) and determine sets of constraint 
equations to be satisfied by the parameters entering the Hamiltonian. Our  results are collected in the 
proposition \ref{prop:results}. In section \ref{sec:sol}, we provide the complete classification of the 
Hamiltonians we can solve by CBA and give in each case the eigenvalues and the eigenfunctions. Finally, 
in section \ref{sec:Hred}, we present simplified versions of the Hamiltonians, including explicit $9\times 9$ 
matricial expressions with physically relevant parameters, and, when possible, connect them with known models.

Associated to this classification, we constructed an interactive web page \cite{web} that can test any 19-vertex Hamiltonian  to determine if it is solvable by the coordinate Bethe ansatz. If so, it provides also the connection with the models we present in this paper, as well as the physical data of the model.

\section{General settings\label{sec:GS}}

\subsection{Hamiltonian\label{sec:Ham}}

We consider a $U(1)$-invariant Hamiltonian $H$ acting on a spin chain of length $L$, where each site carries a $\CC^3$ vector space (i.e. we deal with three-state models).  We assume nearest neighbor interactions, that is 
\beq
H=\sum_{j=1}^L H_{j,j+1},
\eeq
and  periodic conditions, i.e. the site $L+1$ is identified with the site 1. 
As usual in such presentations the indices $(j,j+1)$ indicate where $H_{j,j+1}$ acts non trivially.
The $U(1)$ generator corresponds to the $S^z$ component of the total spin\footnote{Strictly speaking, the spin is $L-S^z$.},
$S^z=\sum_{j=1}^L s^z_j$.
On each site, we choose as basis vectors 
\beq 
|0\rangle = \left(\begin{array}{c} 1\\ 0\\ 0\end{array}\right),\qquad |1\rangle = \left(\begin{array}{c} 0\\ 1\\ 0\end{array}\right),\qquad |2\rangle = \left(\begin{array}{c} 0\\ 0\\ 1\end{array}\right),
\mb{with} s^z|j\rangle =j\,|j\rangle .
\eeq

Under these requirements, the most general two-site Hamiltonian takes the form
\begin{align}
H_{12} = {} & \sum_{i_1,i_2,j_1,j_2 \in \{0,1,2\}} h_{i_1\;i_2}^{j_1\;j_2} \, E_{i_1,j_1} \otimes E_{i_2,j_2} \nonumber \\
= {} & p E_{01} \otimes E_{10} + q E_{10} \otimes E_{01} + t_1 E_{21} \otimes E_{01} + s_1 E_{12} \otimes E_{10} + t_2 E_{01} \otimes E_{21} + s_2 E_{10} \otimes E_{12} \nonumber \\
&+ t_3 E_{12} \otimes E_{21} + s_3 E_{21} \otimes E_{12} + t_p E_{02} \otimes E_{20} + s_p E_{20} \otimes E_{02} + \sum_{i,j} v_{ij} E_{ii} \otimes E_{jj},
\end{align}
where $E_{ij}$ denote the elementary $3 \times 3$ matrices with entry 1 in position $(i,j)$ and zero elsewhere.
In matricial form, it reads
\begin{equation}
H_{12} = 
\begin{pmatrix}
v_{00} & 0 & 0 & 0 & 0 & 0 & 0 & 0 & 0 \\
0 & v_{01} & 0 & p & 0 & 0 & 0 & 0 & 0 \\
0 & 0 & v_{02} & 0 & t_2 & 0 & t_p & 0 & 0 \\
0 & q & 0 & v_{10} & 0 & 0 & 0 & 0 & 0 \\
0 & 0 & s_2 & 0 & v_{11} & 0 & s_1 & 0 & 0 \\
0 & 0 & 0 & 0 & 0 & v_{12} & 0 & t_3 & 0 \\
0 & 0 & s_p & 0 & t_1 & 0 & v_{20} & 0 & 0 \\
0 & 0 & 0 & 0 & 0 & s_3 & 0 & v_{21} & 0 \\
0 & 0 & 0 & 0 & 0 & 0 & 0 & 0 & v_{22} \\
\end{pmatrix}.
\label{eq:ham12}
\end{equation}

We aim at finding the most general Hamiltonian of the form \eqref{eq:ham12} that is solvable by generalized coordinate Bethe ansatz (CBA). This will lead to an exhaustive classification of the possible sets of constraints on the parameters entering $H$, see section \ref{sec:sol}.

Before performing this calculation, we use the symmetries of the problem to keep only physically relevant parameters.

\subsection{Symmetries and transformations\label{sec:sym}}

The Hamiltonian \eqref{eq:ham12} exhibits some symmetries that allow us to simplify it.
\begin{itemize}
\item
\textbf{Telescopic terms.} For any matrix $A$, let us consider the following transformation of the local Hamiltonian:
\begin{equation}
H'_{j,j+1} =  H_{j,j+1} + A_j - A_{j+1}.
\label{eq:telescopic}
\end{equation}
Then the periodicity condition implies that 
\beq
H'= \sum_{j=1}^L H'_{j,j+1} = \sum_{j=1}^L H_{j,j+1}=H.
\eeq
Demanding the $U(1)$ invariance to be preserved forces the matrix $A$ to be diagonal: $A = \diag(a_1,a_2,a_3)$.

The transformation \eqref{eq:telescopic} for diagonal matrix, which involves only two independent parameters, say $a_1-a_2$ and $a_1-a_3$, leads to the following invariant combinations of the parameters: 
\beq
\begin{split}
&V=v_{01}+v_{10}-2v_{00}\,,\quad X_{11}=v_{11}-v_{00}-V\,,\quad Y=v_{02}+v_{20}-2v_{00}-2V\,,\\
& X_{12}=v_{12}+v_{20}-v_{10}-v_{00}-2V\,,\quad
X_{21}=v_{21}+v_{02}-v_{01}-v_{00}-2V\,,\\
& X_{22}=v_{22}-v_{00}-2V.
\end{split}
\eeq
Note that this choice is not unique:
%\footnote{Another possible choice is $V=\half\,(v_{22}-v_{00})$, $X_{11}=v_{11}-\half\,(v_{02}+v_{20})$, $Y=v_{22}+v_{00}+v_{02}+v_{20}-v_{01}-v_{10}-v_{12}-v_{21}$, $X_{12}=v_{12}-v_{10}+\half\,(v_{00}-v_{22}+v_{20}-v_{02})$, $X_{21}=v_{21}-v_{01}+\half\,(v_{00}-v_{22}+v_{02}-v_{20})$, $X_{22}=v_{22}+v_{00}-v_{02}-v_{20}$, which have nice properties under the charge conjugation transformation: $X_{11},Y,X_{22}$ are then invariant and $X_{12},X_{21},V$ are changed in their opposite, compare with eq. \eqref{eq:chargeconj} below.} 
in fact, the combinations above  appear naturally when dealing with the coordinate Bethe ansatz, see section \ref{sec:CBA}.

\item
As already mentioned, we are considering $\mathbf{U(1)}$\textbf{-invariant Hamiltonians.} This implies in particular that the entries of $H_{12}$ satisfy $h_{i_1\;i_2}^{j_1\;j_2} = 0$ if $i_1+i_2 \ne j_1+j_2$, as it can be checked from eq. \eqref{eq:ham12}. In order to get a symmetry of rank one only, one has to impose $(t_1,t_2,s_1,s_2) \ne (0,0,0,0)$, condition that we assume to be satisfied in the whole paper. Indeed, the rank of the symmetry algebra determines the form of the CBA one should use. Hence, it is necessary to fix this rank. The only diagonal generators that commute with the Hamiltonian are then the identity matrix $\II$ and the $S^z$ component of the total spin given in section \ref{sec:Ham}. 
This property can be used to set the zero of the energy for example. A particular interesting choice is to consider $H_{12} - \half\,V (s_1^z+s_2^z)$.

Of course, a further diagonal element can be removed from the Hamiltonian using the identity. In the following we choose $v_{00}=0$.
\end{itemize}

\medskip

One can then consider the following  transformations:

\begin{itemize}
\item
\textbf{Parity transformation (P):} $h_{i_1\;i_2}^{j_1\;j_2} \to h_{i_2\;i_1}^{j_2\;j_1}$ (that is $H_{12} \to H_{21}$), which corresponds to the following correspondence between the parameters ($X_{11}, Y, X_{22}$ are invariant):
\begin{equation}
p\leftrightarrow q, \quad t_1\leftrightarrow t_2, \quad s_1\leftrightarrow s_2, \quad t_3\leftrightarrow s_3, \quad t_p\leftrightarrow s_p, \quad X_{12}\leftrightarrow X_{21} 
\label{eq:parity}
\end{equation}
The Hamiltonians $H_{12}$ and $H_{21}$ lead to systems  where the chain is oriented from right to left instead of left to right.   
Therefore, the set of solutions that lead to solvable Hamiltonian has therefore to be invariant under the correspondence \eqref{eq:parity}.

\item
\textbf{Time reversal (T):} $h_{i_1\;i_2}^{j_1\;j_2} \to h_{j_1\;j_2}^{i_1\;i_2}$ (that is $H_{12} \to H_{12}^t$), which corresponds to the following correspondence between the parameters (all diagonal terms are invariant):
\begin{equation}
p\leftrightarrow q, \quad t_1\leftrightarrow s_1, \quad t_2\leftrightarrow s_2, \quad t_3\leftrightarrow s_3, \quad t_p\leftrightarrow s_p 
\label{eq:timereversal}
\end{equation}

\item
\textbf{Charge conjugation (C):} $h_{i_1\;i_2}^{j_1\;j_2} \to h_{2-i_1\;2-i_2}^{2-j_1\;2-j_2}$ (i.e. indices 0 and 2 are exchanged and index 1 is invariant), which corresponds to the following correspondence between the parameters:
\beq\begin{split}
& p\leftrightarrow s_3, \quad q\leftrightarrow t_3, \quad t_1\leftrightarrow t_2, \quad s_1\leftrightarrow s_2, \quad t_p\leftrightarrow s_p,  \\
& V\leftrightarrow -V-Y-2X_{22}+X_{12}+X_{21}, \quad X_{11}\leftrightarrow X_{11}+Y+X_{22}-X_{12}-X_{21},  \\
& Y+X_{22}\leftrightarrow 5(Y+X_{22})-4(X_{12}+X_{21}), \quad Y-X_{22}\leftrightarrow Y-X_{22},  \\
& X_{12}+X_{21}\leftrightarrow 6(Y+X_{22})-5(X_{12}+X_{21}), \quad X_{12}-X_{21}\leftrightarrow X_{21}-X_{12}
\end{split}
\label{eq:chargeconj}
\eeq
The action of the charge conjugation is equivalent to choose as pseudo-vacuum $|\widetilde\Omega\rangle = \bigotimes_{i=1}^L |2\rangle$ instead of  $|\Omega\rangle = \bigotimes_{i=1}^L |0\rangle$, exchanging the roles of the vectors $|0\rangle$ and $|2\rangle$. 
The solution to the problem obtained thanks to the coordinate Bethe ansatz can be reproduced \textit{mutatis mutandis}, but taking into account the correspondence \eqref{eq:chargeconj}. We will use this property in section \ref{sec:Ham02}.
\end{itemize}
Action of these three transformations on solvable Hamiltonians is displayed in table \ref{table:PCT}, see appendix \ref{sec:table}. In the following we will work modulo these transformations.

\section{Coordinate Bethe ansatz\label{sec:CBA}}

We construct Hamiltonian eigenvectors using a generalization of the original coordinate Bethe ansatz, following the techniques developed in \cite{LS,CRA}.

\subsection{Results}
 Since the $S^z$ component of the total spin commutes with the Hamiltonian, one can decompose the space of states $\cH$ into subspaces with fixed $S^z$-eigenvalue  
$$
\displaystyle \cH=\bigoplus_{M=0}^{2L} \cV_M\,,\qquad S^z\,\vph_M=M\, \vph_M\,,\quad \forall \vph_M\in\cV_M,
$$
and look for eigenvectors of $H$ in a given subspace $\cV_M$. 

For $M=0$, we have a one-dimensional subspace corresponding to a particular eigenvector of the Hamiltonian, called the pseudo-vacuum, defined here as $|\Omega\rangle = \bigotimes_{i=1}^L |0\rangle$. It is easy to see that, 
 since we made the choice $v_{00}=0$, $|\Omega\rangle$ is an eigenvector of $H$ with eigenvalue zero. 

 Then, in $\cV_M$, one considers states with $M$ pseudo-excitations obtained by acting with the raising operator on the pseudo-vacuum. More precisely, an elementary state with $M$ pseudo-excitations is given by
\begin{equation}
\vert x_1,\dots,x_M \rangle = \underbrace{|0\rangle \otimes \dots \otimes |0\rangle}_{x_1-1} \otimes |m_1\rangle \otimes \underbrace{|0\rangle \otimes \dots \otimes |0\rangle}_{x_{m_1+1}-x_{m_1}-1} \otimes |m_2\rangle \otimes \underbrace{|0\rangle \otimes \dots \otimes |0\rangle}_{x_{m_1+m_2+1}-x_{m_1+m_2}-1} \otimes |m_3\rangle \otimes \dots 
\label{eq:elemstate}
\end{equation}
where $1\leq x_1\leq x_2\leq...\leq x_M\leq L$.

The $x_j$'s are the locations of the pseudo-excitations along the chain, and $m_k \in \{1,2\}$ such that $\sum m_k = M$. For $j=1+m_1+\dots+m_{k-1}$, one has $m_k=2$ if $x_{j+1}=x_j$ and $m_k=1$ otherwise. \\
These states form a basis of the subspace ${\cal V}_M$ of states with a given number $M$ of pseudo-excitations.

An eigenstate $\Psi_M$ for the Hamiltonian in a given sector with $M$ pseudo-excitations is obtained as a linear combination of the elementary states \eqref{eq:elemstate} with coefficients $a(x_1,\dots,x_M)$, which are complex-valued functions to be determined:
\begin{equation}\label{eq:psiM}
\Psi_M = \sum_{1 \le x_1 \le \dots \le x_M \le L} a(x_1,\dots,x_M) \vert x_1,\dots,x_M \rangle.
\end{equation}
We assume a plane wave decomposition for the functions $a(x_1,\dots,x_M)$:
\begin{equation}
a(x_1,\dots,x_M) = \sum_{\sigma \in {\mathfrak S}_M} A_\sigma^{(j_1,\dots,j_P)}(k_1,\dots,k_M) \exp\left(\sum_{n=1}^M ik_{\sigma(n)} x_n \right) = \sum_{\sigma \in {\mathfrak S}_M} A_\sigma^{(j_1,\dots,j_P)}(\vec{k}) e^{i\vec{k_\sigma}\cdot\vec{x}}.
\label{eq:planewave}
\end{equation}
Here ${\mathfrak S}_M$ is the permutation group of $M$ elements and $A_\sigma^{(j_1,\dots,j_P)}(k_1,\dots,k_M)$ are functions on the symmetric group algebra depending on some parameters $k$ to be determined later (these are solutions of the so-called Bethe ansatz equations). The indices $(j_1,\dots,j_P)$ correspond to double excitations, i.e. indices such that $x_{j_k+1}=x_{j_k}$ for $k=1,\dots,P$. When there are no double excitations, the
indices $(j_1,\dots,j_P)$ are of course omitted.

\begin{prop}\label{prop:results}
The Hamiltonian $H$ given in \eqref{eq:ham12} is solvable by CBA provided its parameters obey the constraints 
 given in \eqref{E21}, \eqref{E12} and \eqref{E22}. The complete set of solutions to these equations is given in section \ref{sec:sol}, modulo the symmetries mentioned in section \ref{sec:sym} and displayed in table \ref{table:PCT}. 
 
For CBA-solvable Hamiltonians $H$, the state \eqref{eq:psiM} is an eigenvector of  $H$, with energy
\begin{equation}
E_M = MV + \sum_{n=1}^M (q\,e^{ik_{n}}+p\,e^{-ik_{n}}) 
\label{eq:energie}
\end{equation}
provided the coefficients $A_\sigma$ and $A_\sigma^{(j_1,\dots,j_P)}$, $P=1,...,\left[\frac{M}2\right]$,
are related by
\begin{align}
&\frac{A_{\sigma T_j}(\vec{k})}{A_{\sigma}(\vec{k})} \equiv S(k_{\sigma(j)},k_{\sigma(j+1)}) = -\frac{\Lambda(k_{\sigma(j)},k_{\sigma(j+1)})}{\Lambda(k_{\sigma(j+1)},k_{\sigma(j)})}\,,
\label{eq:Smatrix}
\\
&\frac{A_{\sigma}^{(j)}(\vec{k})}{A_{\sigma}(\vec{k})} \equiv N(k_{\sigma(j)},k_{\sigma(j+1)}) = \frac{(e^{ik_{\sigma(j)}}-e^{ik_{\sigma(j+1)}})(p+qe^{ik_{\sigma(j)}+ik_{\sigma(j+1)}})(t_2+t_1e^{ik_{\sigma(j)}+ik_{\sigma(j+1)}})}{2\,\Lambda(k_{\sigma(j+1)},k_{\sigma(j)})},
\label{eq:Nfactor}\\
&\frac{A_{\sigma}^{(j_1,...,j_P)}(\vec{k})}{A_{\sigma}(\vec{k})} = \prod_{n=1}^P N(k_{\sigma(j_n)},k_{\sigma(j_n+1)}),
\label{eq:Nmultiple}
\end{align}
where $T_j \in {\mathfrak S}_M$ denotes the transposition $(j,j+1)$ and
\begin{align}
&\Lambda(k_{j},k_{n}) = {}    e^{ik_{n}}(s_1+s_2e^{ik_{j}+ik_{n}})(t_2+t_1e^{ik_{j}+ik_{n}}) \\
&-\Big( Y e^{ik_{j}+ik_{n}} - qe^{ik_{j}+ik_{n}}(e^{ik_{j}}+e^{ik_{n}}) - p(e^{ik_{j}}+e^{ik_{n}})
 + s_p e^{2ik_{j}+2ik_{n}} + t_p \Big) \big( X_{11} e^{ik_{n}} - qe^{ik_{j}+ik_{n}} -p \big).
\nonumber
\end{align}
The momenta $k_j$ must also obey the Bethe equations
\begin{equation}
e^{ik_{j}L} = \prod_{n \ne j} S(k_n,k_j)\,, \qquad j=1,...,M.
\label{eq:BAE}
\end{equation}
\end{prop}
Remark that when  $p=q=0$,  the energy depends only on the number of pseudo-excitations. 
In this case, one needs to consider another vacuum to get a complete spectrum, see section \ref{sec:Ham02}.
When $p$ and $q$ are both non vanishing, the energy can be rewritten as
\begin{equation}
E_M = MV + \sqrt{pq} \sum_{n=1}^M (z_n\sqrt{\theta} + \frac{1}{z_n\sqrt{\theta}}) \quad \text{where} 
\quad \theta=q/p \mb{and} z_n=e^{ik_n}.
\end{equation}
In this case, after eliminating the constant term $MV$ thanks to the $S_z$ operator and rescaling 
of the Hamiltonian by $\sqrt{pq}$, the energy clearly depends only on the parameter $\theta$ (and those of the S-matrix through the Bethe equations).

\subsection{Proofs}

Since $H$ is a sum of two-site operators $H_{j,j+1}$, one has to single out only the following configurations:

\medskip

\begin{enumerate}
\item{\textbf{Configurations leading to the determination of eigenvalues and eigenvectors:}} \\
$\centerdot$ the $x_j$'s are far from each other (``generic case''), \\
$\centerdot$ $x_{j+1}=x_j+1$ for some $j$ and the other $x_n$'s are far from each other, \\
$\centerdot$ $x_{j+1}=x_j$ for some $j$ and the other $x_n$'s are far from each other, \\
$\centerdot$ $x_{j_k+1}=x_{j_k}$, $x_{j_k}+1 < x_{j_k+2}$ for $k=1,...,P$, and the other $x_n$'s are far from each other, 
\item{\textbf{Configurations leading to constraints on the parameters of the models:}} \\
$\centerdot$ $x_{j+1}=x_j$ and $x_{j+2}=x_{j}+1$ for a given $j$ , the other $x_n$'s are far from each other, \\
$\centerdot$ $x_{j+1}=x_j$ and $x_{j-1}=x_{j}-1$ for a given $j$, the other $x_n$'s are far from each other, \\
$\centerdot$ $x_{j-1}=x_{j}$ and $x_{j+1}=x_{j+2}=x_{j}+1$, the other $x_n$'s are far from each other.
\item{\textbf{Configurations leading to Bethe equations and/or already known eqs:}} \\
$\centerdot$ $x_1=1$ and the other $x_n$'s are far from each other, \\
$\centerdot$ $x_M=L$ and the other $x_n$'s are far from each other, \\
$\centerdot$ $x_1=1$, $x_M=L$ and the other $x_n$'s are far from each other, \\
$\centerdot$ $x_1=x_2=1$ (or equivalently $x_{M-1}=x_M=L$) and the other $x_n$'s are far from each other, \\
$\centerdot$ $x_1=x_2=1$ and $x_M=L$ (or equivalently $x_1=1$ and $x_{M-1}=x_M=L$), and the other $x_n$'s are far from each other, \\
$\centerdot$ $x_1=x_2=1$ and $x_{M-1}=x_M=L$, and the other $x_n$'s are far from each other.
\end{enumerate}

\subsubsection{Configurations leading to energy and eigenstates}

$\bullet$ \textbf{Configuration where the $x_j$'s are generic}, i.e. are
far from each other and from the edges: $1 < x_1 \ll ... \ll x_n \ll x_{n+1} \ll ... \ll x_M < L$. Projecting the Schr\"odinger equation on it, we get
\begin{equation}
\sum_{\sigma \in {\mathfrak S}_M} A_\sigma(\vec{k}) e^{i\vec{k_\sigma}\cdot\vec{x}} \left( MV + \sum_{n=1}^M (q e^{ik_{\sigma(n)}} + p e^{-ik_{\sigma(n)}}) \right) = E_M \sum_{\sigma \in {\mathfrak S}_M} A_\sigma(\vec{k}) e^{i\vec{k_\sigma}\cdot\vec{x}} 
\end{equation}
which leads to the value \eqref{eq:energie} of the energy of the state $\Psi_M$.

\medskip

$\bullet$ \textbf{Configuration where $x_{j+1}=x_j+1$ for a given $j$ (not on the edges)}, the other $x_n$'s being far from each other and from the edges. Then one gets
\begin{equation}
\sum_{\sigma \in {\mathfrak S}_M} e^{i\vec{k_\sigma}\cdot\vec{x}} \left( A_\sigma(\vec{k}) \left( X_{11} - q e^{ik_{\sigma(j)}} - p e^{-ik_{\sigma(j+1)}} \right) + A_\sigma^{(j)}(\vec{k}) \left( s_2 e^{ik_{\sigma(j)}} + s_1 e^{-ik_{\sigma(j+1)}} \right) \right) = 0
\end{equation}
Note that, since $x_{j+1}=x_j+1$, one has  here
$$\displaystyle e^{i\vec{k_\sigma}\cdot\vec{x}} = e^{ik_{\sigma(j+1)}} \exp\left( i(k_{\sigma(j)}+k_{\sigma(j+1)})x_j+\sum_{n \ne j,j+1} ik_{\sigma(n)} x_n   \right),
$$
 which implies a symmetrization in the exchange $j \leftrightarrow j+1$ before projecting onto the independent states \eqref{eq:elemstate}. Hence one gets, where $T_j \in {\mathfrak S}_M$ denotes the transposition $(j,j+1)$,
\begin{align}
&A_{\sigma}(\vec{k}) e^{ik_{\sigma(j+1)}} \left( X_{11} - q e^{ik_{\sigma(j)}} - p e^{-ik_{\sigma(j+1)}} \right) + 
 A_{\sigma T_j}(\vec{k}) e^{ik_{\sigma(j)}} \left( X_{11} - q e^{ik_{\sigma(j+1)}} - p e^{-ik_{\sigma(j)}} \right) 
 + \nonumber \\
& \left(A_{\sigma T_j}^{(j)}(\vec{k}) + A_{\sigma}^{(j)}(\vec{k}) \right)\left( s_2 e^{ik_{\sigma(j+1)+ik_{\sigma(j)}}} + s_1  \right) 
= 0.
\label{eq:constrmatS1}
\end{align}

$\bullet$ \textbf{Configuration where $x_{j+1}=x_j$ for a given $j$ (not on the edges)}, the other $x_n$'s being far from each other and from the edges. Then one has
\begin{align}
\sum_{\sigma \in {\mathfrak S}_M} e^{i\vec{k_\sigma}\cdot\vec{x}} & \Big( A_\sigma^{(j)}(\vec{k}) \big( Y - q e^{ik_{\sigma(j)}} - q e^{ik_{\sigma(j+1)}} - p e^{-ik_{\sigma(j)}} - p e^{-ik_{\sigma(j+1)}} \nonumber \\
& + s_p e^{ik_{\sigma(j)}+ik_{\sigma(j+1)}} + t_p e^{-ik_{\sigma(j)}-ik_{\sigma(j+1)}} \big) + A_\sigma(\vec{k}) \big( t_1 e^{ik_{\sigma(j+1)}} + t_2 e^{-ik_{\sigma(j)}} \big) \Big) = 0.
\end{align}
After symmetrization in $(j,j+1)$ as above, one obtains
\begin{align}
& \big( A_{\sigma}^{(j)}(\vec{k}) + A_{\sigma T_j}^{(j)}(\vec{k}) \big) \big( Y - q e^{ik_{\sigma(j)}} - q e^{ik_{\sigma(j+1)}} - p e^{-ik_{\sigma(j)}} - p e^{-ik_{\sigma(j+1)}} + s_p e^{ik_{\sigma(j)}+ik_{\sigma(j+1)}} \nonumber \\
& + t_p e^{-ik_{\sigma(j)}-ik_{\sigma(j+1)}} \big) + A_{\sigma}(\vec{k}) \big( t_1 e^{ik_{\sigma(j+1)}} + t_2 e^{-ik_{\sigma(j)}} \big) + 
A_{\sigma T_j}(\vec{k}) \big( t_1 e^{ik_{\sigma(j)}} + t_2 e^{-ik_{\sigma(j+1)}} \big) = 0.
\label{eq:constrmatS2}
\end{align}
Without any loss of generality, we choose\footnote{Another possible choice \cite{AB} is to impose $A_{\sigma T_j}^{(j)}(\vec{k}) = S(k_{\sigma(j)},k_{\sigma(j+1)})\,A_{\sigma}^{(j)}(\vec{k})$. One goes from one choice to another through the renormalisation $A_{\sigma}^{(j)}(\vec{k})\ \to\ (k_{\sigma(j)}-k_{\sigma(j+1)})\Lambda(k_{\sigma(j)},k_{\sigma(j+1)})\,A_{\sigma}^{(j)}(\vec{k})$.} to impose $A_{\sigma T_j}^{(j)}(\vec{k}) = A_{\sigma}^{(j)}(\vec{k})$. Then, using eqs. \eqref{eq:constrmatS1} and \eqref{eq:constrmatS2}, one gets  relations \eqref{eq:Smatrix} and \eqref{eq:Nfactor}.

\medskip

$\bullet$ \textbf{Configuration where $x_{j_k+1}=x_{j_k}$ for $k=1,...,P$} and the other $x_n$'s being far from each other and from the edges. One gets
\begin{align}
& \sum_{\sigma \in {\mathfrak S}_M} e^{i\vec{k_\sigma}\cdot\vec{x}} \; \bigg\{ A_\sigma^{(j_1,...,j_P)}(\vec{k}) \Big( PY  + \sum_{k=1}^P t_p e^{-ik_{\sigma(j_k)}-ik_{\sigma(j_k+1)}} - p e^{-ik_{\sigma(j_k)}} - p e^{-ik_{\sigma(j_k+1)}} \nonumber \\
& + s_p e^{ik_{\sigma(j_k)}+ik_{\sigma(j_k+1)}} - q e^{ik_{\sigma(j_k)}} - q e^{ik_{\sigma(j_k+1)}} \Big) + \sum_{k=1}^P A_\sigma^{(j_1...\check{\jmath}_k...j_P)}(\vec{k}) ( t_2 e^{-ik_{\sigma(j_k)}} + t_1 e^{ik_{\sigma(j_k+1)}} ) \bigg\} = 0
\end{align}
where $A_\sigma^{(j_1...\check{\jmath}_k...j_P)}(\vec{k})$ means that $x_{j_n+1}=x_{j_n}$ for $n=1,...,P$ and $n \ne k$. \\
Morevover, after projection onto the states \eqref{eq:elemstate}, one needs to symmetrize (independently) on each pair $(j_n,j_n+1)$. One is led to a recursion relation linking $A_\sigma^{(j_1,...,j_P)}(\vec{k})$ and $A_\sigma^{(j_1,...,j_{P-1})}(\vec{k})$ that can be solved, and one gets \eqref{eq:Nmultiple}.

\subsubsection{Configurations leading to constraints on parameters}

$\bullet$ \textbf{Configuration where $x_{j+1}=x_j$ and $x_{j+2}=x_{j}+1$ for a given $j$ (not on the edges)}, the other $x_n$'s being far from each other and from the edges. One obtains
\begin{align}
\sum_{\sigma \in {\mathfrak S}_M} e^{i\vec{k_\sigma}\cdot\vec{x}} & \Big( A_\sigma^{(j)}(\vec{k}) \big( X_{21} - q e^{ik_{\sigma(j)}} - q e^{ik_{\sigma(j+1)}} - p e^{-ik_{\sigma(j)}} - p e^{-ik_{\sigma(j+1)}} - p e^{-ik_{\sigma(j+2)}} \nonumber \\
& + t_p e^{-ik_{\sigma(j)}-ik_{\sigma(j+1)}} \big) + A_\sigma^{(j+1)}(\vec{k}) s_3 e^{ik_{\sigma(j+1)}} + A_\sigma(\vec{k}) t_2 e^{-ik_{\sigma(j)}} \Big) = 0.
\end{align} 
Here one has $\displaystyle e^{i\vec{k_\sigma}\cdot\vec{x}} = e^{ik_{\sigma(j+2)}} \exp\bigg( \sum_{n \ne j,j+1,j+2} ik_{\sigma(n)} x_n + \sum_{n = j,j+1,j+2} ik_{\sigma(n)} x_j \bigg)$ given the configuration. Therefore, projecting onto the states \eqref{eq:elemstate}, it is now necessary to symmetrize on $(j,j+1,j+2)$. Taking into account the relations \eqref{eq:Smatrix} and \eqref{eq:Nfactor} that allow one to express all $A$ functions in terms of $A_\sigma(\vec{k})$ only, one gets now 
\beq\label{E21}
\sum_{\sigma \in {\mathfrak S}_3} {\cal E}_{21}(k_{\sigma(j)},k_{\sigma(j+1)},k_{\sigma(j+2)}) = 0, 
\eeq
where
\begin{align}
& {\cal E}_{21}(k_{\sigma(j)},k_{\sigma(j+1)},k_{\sigma(j+2)}) = A_\sigma(\vec{k}) \, e^{ik_{\sigma(j+2)}} \Big( N(k_{\sigma(j)},k_{\sigma(j+1)}) \big( X_{21} - q e^{ik_{\sigma(j)}} - q e^{ik_{\sigma(j+1)}} - p e^{-ik_{\sigma(j)}} \nonumber \\
& - p e^{-ik_{\sigma(j+1)}} - p e^{-ik_{\sigma(j+2)}} + t_p e^{-ik_{\sigma(j)}-ik_{\sigma(j+1)}} \big) + N(k_{\sigma(j+1)},k_{\sigma(j+2)}) \, s_3 e^{ik_{\sigma(j+1)}} + t_2 e^{-ik_{\sigma(j)}} \Big).
\end{align}
Then projecting the above constraint onto the monomials in the variables 
$e^{ik_{\sigma(\ell)}}$, $\ell=j,\,j+1,\,j+2$, one gets a first set of 32 constraint equations.
For sake of simplicity, we avoid writing these equations here.

\medskip

$\bullet$ \textbf{Configuration where $x_{j+1}=x_j$ and $x_{j-1}=x_{j}-1$ for a given $j$ (not on the edges)}, the other $x_n$'s being far from each other and from the edges. In the same way, when one considers such a configuration, one obtains 
\begin{align}
\sum_{\sigma \in {\mathfrak S}_M} e^{i\vec{k_\sigma}\cdot\vec{x}} & \Big( A_\sigma^{(j)}(\vec{k}) \big( X_{12} - q e^{ik_{\sigma(j-1)}} - q e^{ik_{\sigma(j)}} - q e^{ik_{\sigma(j+1)}} - p e^{-ik_{\sigma(j)}} - p e^{-ik_{\sigma(j+1)}} \nonumber \\
& + s_p e^{ik_{\sigma(j)}+ik_{\sigma(j+1)}} \big) + A_\sigma^{(j-1)}(\vec{k}) \, t_3 e^{-ik_{\sigma(j)}} + A_\sigma(\vec{k}) \, t_1 e^{ik_{\sigma(j+1)}} \Big) = 0.
\end{align}
Again, after projection onto the states \eqref{eq:elemstate}, one is left to symmetrize on $(j-1,j,j+1)$, 
\beq\label{E12}
\sum_{\sigma\in{\mathfrak S}_3}  \cE_{12}(k_{\sigma(j-1)},k_{\sigma(j)},k_{\sigma(j+1)})=0
\eeq
where  the expression $ \cE_{12}$
is  given by
\begin{align}
& \cE_{12}(k_{\sigma(j-1)},k_{\sigma(j)},k_{\sigma(j+1)}) = e^{-ik_{\sigma(j)}} \Big( N(k_{\sigma(j)},k_{\sigma(j+1)}) \big( X_{12} - q e^{ik_{\sigma(j-1)}} - q e^{ik_{\sigma(j)}} - q e^{ik_{\sigma(j+1)}} \nonumber \\
& - p e^{-ik_{\sigma(j)}} - p e^{-ik_{\sigma(j+1)}} + s_p e^{ik_{\sigma(j)}+ik_{\sigma(j+1)}} \big) + N(k_{\sigma(j-1)},k_{\sigma(j)}) t_3 e^{-ik_{\sigma(j)}} + t_1 e^{ik_{\sigma(j+1)}} \Big).
\end{align}

The projection of the constraint equation onto the monomials in the variables $e^{ik_{\sigma(\ell)}}$, $\ell=j-1,\,j,\,j+1$ leads to a second set of 32 constraint equations. 

\medskip

$\bullet$ \textbf{Configuration where $x_{j-1}=x_{j}$ and $x_{j+1}=x_{j+2}=x_{j}+1$}, the other $x_n$'s being far from each other and from the edges. One gets
\begin{align}
\sum_{\sigma \in {\mathfrak S}_M} e^{i\vec{k_\sigma}\cdot\vec{x}} & \Big( A_\sigma^{(j-1,j+1)}(\vec{k}) \big( X_{22} + t_p e^{-ik_{\sigma(j-1)}-ik_{\sigma(j)}} - q e^{ik_{\sigma(j-1)}} - q e^{ik_{\sigma(j)}} - q e^{ik_{\sigma(j+1)}} - q e^{ik_{\sigma(j+2)}} \nonumber \\
& + s_p e^{ik_{\sigma(j+1)}+ik_{\sigma(j+2)}} - p e^{-ik_{\sigma(j-1)}} - p e^{-ik_{\sigma(j)}}  - p e^{-ik_{\sigma(j+1)}}  - p e^{-ik_{\sigma(j+2)}} \big) \nonumber \\
& + A_\sigma^{(j+1)}(\vec{k}) \, t_2 e^{-ik_{\sigma(j-1)}} + A_\sigma^{(j-1)}(\vec{k}) \, t_1 e^{ik_{\sigma(j+2)}} \Big) = 0.
\end{align}
Since now $\displaystyle e^{i\vec{k_\sigma}\cdot\vec{x}} = e^{ik_{\sigma(j+1)}+ik_{\sigma(j+2)}} \exp\bigg( \sum_{n \ne j-1,j,j+1,j+2} ik_{\sigma(n)} x_n + \sum_{n = j-1,j,j+1,j+2} ik_{\sigma(n)} x_j \bigg)$, one symmetrizes on $(j-1,j,j+1,j+2)$, and  gets 
\beq\label{E22}
\sum_{\sigma \in {\mathfrak S}_4} {\cal E}_{22}(k_{\sigma(j-1)},k_{\sigma(j)},k_{\sigma(j+1)},k_{\sigma(j+2)}) = 0,
\eeq
 where
\begin{align}
& {\cal E}_{22} = A_\sigma(\vec{k}) \, e^{ik_{\sigma(j+1)}+ik_{\sigma(j+2)}} \bigg( N(k_{\sigma(j-1)},k_{\sigma(j)}) N(k_{\sigma(j+1)},k_{\sigma(j+2)}) \big( X_{22} + t_p e^{-ik_{\sigma(j-1)}-ik_{\sigma(j)}} \nonumber \\
& - q e^{ik_{\sigma(j-1)}} - q e^{ik_{\sigma(j)}} - q e^{ik_{\sigma(j+1)}} - q e^{ik_{\sigma(j+2)}} + s_p e^{ik_{\sigma(j+1)}+ik_{\sigma(j+2)}} - p e^{-ik_{\sigma(j-1)}} - p e^{-ik_{\sigma(j)}} \nonumber \\
& - p e^{-ik_{\sigma(j+1)}} - p e^{-ik_{\sigma(j+2)}} \big) + N(k_{\sigma(j+1)},k_{\sigma(j+2)}) \, t_2 e^{-ik_{\sigma(j-1)}} + N(k_{\sigma(j-1)},k_{\sigma(j)}) \, t_1 e^{ik_{\sigma(j+2)}} \bigg).
\end{align}
The projection of the constraint equation onto the monomials in the variables $e^{ik_{\sigma(\ell)}}$, 
$\ell=j-1,\,j,\,j+1,\,j+2$ finally leads to a third set of constraint equations.

\medskip

The solutions to the sets of equations \eqref{E21}, \eqref{E12} and \eqref{E22} give necessary conditions to be satisfied among the parameters defining the two-site Hamiltonian \eqref{eq:ham12} to ensure the solvability of the chain. This leads to a classification of three-state integrable models as shown in the next section.

\subsubsection{Configurations leading to the Bethe equations}

We now concentrate on configurations with at least one excitation lying on the chain edges 1 and/or $L$. Using the periodicity condition of the chain, this will allow us to derive the equations that determine the admissible values of the parameters $k$ entering into the definition of the plane wave \eqref{eq:planewave}, i.e. the Bethe ansatz equations.

\medskip

$\bullet$ \textbf{Configuration where $x_1=1$} and the other $x_n$'s are far from each other and from the edges: $1 = x_1 \ll ... \ll x_n \ll x_{n+1} \ll ... \ll x_M < L$. Then one gets provided that $p \ne 0$,
\begin{equation}
\sum_{\sigma \in {\mathfrak S}_M} A_\sigma(\vec{k}) \left( \exp \Big(\sum_{n=2}^M ik_{\sigma(n)} x_n \Big) - \exp \Big(ik_{\sigma(M)} L + \sum_{n=2}^M ik_{\sigma(n-1)} x_n \Big) \right) = 0.
\end{equation}
Performing the transformation $\sigma \to \sigma T_1 \dots T_{M-1}$ in the second term, one gets $A_{\sigma T_1 \dots T_{M-1}}(\vec{k}) = e^{-ik_{\sigma(1)}L} A_\sigma(\vec{k})$. Taking into account \eqref{eq:Smatrix}, we finally obtain the Bethe ansatz equations \eqref{eq:BAE}.

In the same way, one can consider a configuration where $x_M=L$ and the other $x_n$'s are far from each other and from the edges: $1 < x_1 \ll ... \ll x_n \ll x_{n+1} \ll ... \ll x_M = L$. Then one gets provided that $q \ne 0$,
\begin{equation}
\sum_{\sigma \in {\mathfrak S}_M} A_\sigma(\vec{k}) \left( \exp \Big(ik_{\sigma(M)} (L+1) + \sum_{n=1}^{M-1} ik_{\sigma(n)} x_n \Big) - \exp \Big(ik_{\sigma(1)} + \sum_{n=1}^{M-1} ik_{\sigma(n+1)} x_n \Big) \right) = 0.
\end{equation}
Now, performing the transformation $\sigma \to \sigma T_{M-1} \dots T_1$ in the second term, one gets 
$$A_{\sigma T_{M-1} \dots T_1}(\vec{k}) = e^{ik_{\sigma(M)}L} A_\sigma(\vec{k}),$$ which leads also to equation \eqref{eq:BAE}. 

Note that since we excluded the values $p=q=0$, the BAE \eqref{eq:BAE} holds in any case.

\medskip

$\bullet$ \textbf{Configuration where $x_1=1$, $x_M=L$} and the other $x_n$'s are far from each other: $1 = x_1 \ll ... \ll x_n \ll x_{n+1} \ll ... \ll x_M = L$. One obtains
\begin{align}
\sum_{\sigma \in {\mathfrak S}_M} & A_\sigma(\vec{k}) \exp \Big(ik_{\sigma(1)} + ik_{\sigma(M)} L + \sum_{n=2}^{M-1} ik_{\sigma(n)} x_n \Big) \big( X_{11} - q e^{ik_{\sigma(M)}} - p e^{-ik_{\sigma(1)}} \big) \nonumber \\
& + A_\sigma^{(M-1)}(\vec{k}) \exp \Big(ik_{\sigma(M-1)} L + ik_{\sigma(M)} L + \sum_{n=2}^{M-1} ik_{\sigma(n-1)} x_n \Big) \nonumber \\
& + A_\sigma^{(1)}(\vec{k}) \exp \Big(ik_{\sigma(1)} + ik_{\sigma(2)} + \sum_{n=2}^{M-1} ik_{\sigma(n+1)} x_n \Big) = 0.
\end{align}
One then performs the transformations $\sigma \to \sigma T_{1} \dots T_{M-1}$ (second term) and $\sigma \to \sigma T_{M-1} \dots T_1$ (third term) and uses the relations \eqref{eq:Smatrix} and \eqref{eq:Nfactor}. After the necessary symmetrization on the pair $(1,M)$ and projection onto the states \eqref{eq:elemstate}, one is left with an equation expressed in terms of $A_\sigma(\vec{k})$ only, the $S$-matrix and the decay coefficient $N$. Plugging the BAE \eqref{eq:BAE} into the obtained equation, it appears that no further condition is required.

\medskip

$\bullet$ \textbf{Other ``edge configurations''.} They correspond to the following cases: 
\begin{enumerate}[(i)]
\item $x_1=x_2=1$ (or equivalently $x_{M-1}=x_M=L$), 
\item $x_1=x_2=1$ and $x_M=L$ (or equivalently $x_1=1$ and $x_{M-1}=x_M=L$), 
\item $x_1=x_2=1$ and $x_{M-1}=x_M=L$, and the other $x_n$'s are far from each other. 
\end{enumerate}
The approach is similar to the previous case. After obtaining the Schr\"odinger equation for the particular configuration under consideration using the periodicity conditions, one performs the suitable transformations on the permutations and write all functions $A(\vec{k})$ in terms of the running $A_\sigma(\vec{k})$ only, products of $S$-matrices and decay coefficients $N$. If necessary, symmetrize on the indices which are left after the projection on the elementary states \eqref{eq:elemstate}. In each case, plugging the BAE \eqref{eq:BAE} into the equation that is finally obtained leads to some constraint equation belonging to one of the sets \eqref{E21}, \eqref{E12}  or 
\eqref{E22}. No further condition is eventually needed.

\section{Solutions of the constraint equations\label{sec:sol}}

In this section, we present all the non trivial solutions to equations \eqref{E21}, \eqref{E12} and 
\eqref{E22}, described in section \ref{sec:CBA}. It provides a classification of three-state models solvable by CBA. We used a formal calculation software to solve completely these equations, and found 4 nineteen-vertex, 4 seventeen-vertex and 2 fourteen-vertex models, up to the transformations under parity, time reversal and charge conjugation, see section \ref{sec:sym}.  If one includes the images of the irreducible solutions under these transformations, one gets 22 solutions, see table \ref{table:PCT}. 
\begin{rmk}
Of course, when directly solving the equations \eqref{E21}, \eqref{E12} and 
\eqref{E22}, one finds many more solutions, but most of them are subcases of these 10 ``irreducible'' solutions. We developed a sofware that, starting from any given Halmiltonian of the form \eqref{eq:ham12}, can analyze whether the Hamiltonian is solvable by CBA, and, if so, to which one of the 10 irreducible solutions it corresponds. This program is freely accessible on our web page \cite{web}. Let us stress that  the correspondence may be sensitive to  the  choice of free parameters that is used. This is taken into account by the software. However, in this article we made (arbitrarily) one specific choice. The other ones are found through  the image  under parity, charge conjugation and/or time reversal transformations of the choice we present here. We illustrate this in a particular case, see section \ref{sec:sol24}.
\end{rmk}

\textit{In the following, we  classify the models  that have $(t_1,t_2,s_1,s_2)\neq(0,0,0,0)$ and $(p,q,t_3,s_3)\neq(0,0,0,0)$. Because we work modulo $P$, $C$, $T$ transformation, it is enough to present the solutions 
with $(p,q)\neq(0,0)$ and $(t_1,t_2)\neq(0,0)$:} \\
 $(i)$ since we are considering $U(1)$ invariant models, to get a symmetry of rank 1 only (not rank 2), we are led to $(t_1,t_2,s_1,s_2)\neq(0,0,0,0)$. Now suppose that we get a solution with $(t_1,t_2)=(0,0)$. Then, this solution has $(s_1,s_2)\neq(0,0)$. But the image of this solution under time reversal is also a solution and has $(t_1,t_2)\neq(0,0)$ and $(s_1,s_2)=(0,0)$: since we are working modulo this transformation, we can choose to present solutions with $(t_1,t_2)\neq(0,0)$; \\
$(ii)$ to be able to construct the CBA on the vacuum $|\Omega\rangle$ or $|\tilde\Omega\rangle$, one needs to have  
$(p,q,t_3,s_3)\neq(0,0,0,0)$. Now, since charge conjugation \eqref{eq:chargeconj} exchanges $(p,q)$ and $(t_3,s_3)$, we can suppose
 $(p,q)\neq(0,0)$.

These requirements exclude all the cases obtained in \cite{ITA}, but models 7 and 10: the remaining ones are models with rank 2 symmetry, or diagonal Hamiltonians, or not solvable through CBA. 
 They exclude also the model based on Temperley-Lieb algebra \cite{TL}, for which another type of CBA is needed \cite{CBA-TL}.

\medskip
We  introduce the following reduced parameters: 
\beq
\tau_p = t_p/p\,,\quad \tau_2 = t_2/p\,,\quad \tau_3 = t_3/p\,,\quad \theta=q/p\,,\quad \Upsilon = Y/p\,,\quad
 \sigma = s_1t_2/p^2\,,\quad \mu = t_1/t_2.
\eeq
These reduced parameters are the only ones that are part of the physical data of the models: scattering matrix $S$, decay coefficient $N$, energy $E$ and BAEs. Hence the other ones can be eliminated from the model through gauge transformations and/or telescoping terms, as it is done in section \ref{sec:Hred}. We chose to present here our ``raw'' Hamiltonians to be easily compared with any given Hamiltonian. 

These ``raw'' Hamiltonians are defined whatever the values of the free parameters, provided they lead to well-defined expressions for the other parameters. The reduced parameters are valid for generic values of the free parameters and can be ill-defined for some specific values, see remark \ref{rmk:zero} in section \ref{sec:Hred}.

We define $J$ as one solution of the equation $J^2+J+1=0$. 

\subsection{Nineteen vertices\label{sec:19V}}

In this subsection, we focus on the solutions for which all off-diagonal parameters entering in the Hamiltonian are non zero.

\begin{rmk}\label{rmk:19v}
We will call the corresponding Hamiltonian a ``19-vertex'' one, since we get 19 non-vanishing entries for $H_{12}$ when adding the 9 diagonal parameters to the 10 off-diagonal ones. Note however that one can always cancel some of the diagonal entries using the symmetries as discussed in section \ref{sec:sym}.

We would like to stress also that this name 19-vertex is not related to the terminology used for $R$-matrix formalism.
\end{rmk}

\subsubsection{Generalized Zamolodchikov-Fateev model (gZF)\label{sec:sol19}} %19

The parameters which are left free are $p, t_p, t_2, s_1$. The remaining parameters entering the off-diagonal part of the Hamiltonian are given by
\beq
q=s_3=\frac{p^3}{t_p^2}\,,\quad  t_1=\frac{p^2 t_2}{t_p^2}\,,\quad t_3=p\,,\quad s_2=\frac{p^2 s_1}{t_p^2}
\,,\quad s_p=\frac{p^4}{t_p^3},\label{eq:sol19-1}
\eeq
while on the diagonal we get:
\beq
X_{11}=0\,,\quad Y=\frac{2p^2}{t_p}\,,\quad  X_{12}=X_{21}=\frac{3p^2-s_1 t_2}{t_p}\,,\quad  X_{22}= \frac{4p^2-2s_1 t_2}{t_p}.
\label{eq:sol19-2}
\eeq
The S-matrix depends only on the reduced parameters $\tau_p$ and $\sigma$:
\begin{equation}
S(z_1,z_2) = -\frac{z_1z_2 - \tau_p(z_1+z_2-\sigma z_2) + \tau_p^2}{z_1z_2 - \tau_p(z_1+z_2-\sigma z_1) + \tau_p^2} \\
\end{equation}
and the decay coefficient $N$ reads
\begin{equation}
N(z_1,z_2) = \frac{\tau_2\tau_p(z_1-z_2)}{2(z_1z_2 - \tau_p(z_1+z_2-\sigma z_1) + \tau_p^2)}.
\end{equation}

\begin{rmk}
The PT-invariant models of ref. \cite{MP}, branch 1A, are obtained as particular cases of this one. More precisely, setting 
\beq
p=\frac{2k^2}{k^4-1} \,,\quad t_p=\frac{-2\eps_1k^2}{k^4-1} \,,\quad t_2 = s_1 = \pm e^{-\frac{i\pi}{4}(1-\eps_1)} \frac{2k}{k^2-1} \,,
\eeq
one recovers the branch 1A Hamiltonians $H^{\pm}_{1A}(k,\eps_1)$ of ref. \cite{MP} which is associated to the Zamolodchikov-Fateev model \cite{ZF}.
\end{rmk}

\subsubsection{Generalized Izergin-Korepin model (gIK)\label{sec:sol27}} %27

The parameters which are left free are $p, t_p, t_2$. The remaining parameters entering the off-diagonal part of the Hamiltonian are given by
\begin{align}
& s_p=v^4\,\frac{p^4}{t_p^3}\,,\quad \displaystyle s_3=q=v^2\,\frac{p^3}{t_p^2}\,,\quad t_3=p\,,\quad 
t_1=u_{\pm}^{-1} \,\frac{p^2 t_2}{t_p^2}\\
&\displaystyle s_1=v(v-1)\,\frac{p^2}{t_2}\,,\quad \displaystyle s_2=u_{\mp}^{-1}v(v-1) \frac{p^4}{t_2t_p^2} ,
\end{align}
 while on the diagonal we get:
\begin{align}
&X_{11}=v(v+1)\,\frac{p^2}{t_p}\,,\quad \displaystyle Y=(v^2+1)\,\frac{p^2}{t_p}\,,\quad  X_{22}=2(v+1)\,\frac{p^2}{t_p},\\
&\displaystyle X_{12}=(v^2+1-u_{\mp}^{-1})\,\frac{p^2}{t_p}\,,\quad \displaystyle X_{21}=(v^2+1-u_{\pm}^{-1})\,\frac{p^2}{t_p}\,,\end{align}
where $v$ is a free parameter and $u_{\pm}$ are the two solutions of the equation 
\beq
v^4\,Z^2+(1+2v-v^2)Z+1=0.
\eeq
The S-matrix depends only on the reduced parameter $\tau_p$ and $v$:
\begin{equation}
S(z_1,z_2) = -\frac{(v^2 z_1z_2 - \tau_p(z_1+vz_2) + \tau_p^2)(v^2 z_1z_2 - \tau_p(1+v)z_2 + \tau_p^2)}{(v^2 z_1z_2 - \tau_p(z_2+vz_1) + \tau_p^2)(v^2 z_1z_2 - \tau_p(1+v)z_1 + \tau_p^2)}
\end{equation}
and the decay coefficient $N$ reads
\begin{equation}
N(z_1,z_2) = \frac{\tau_2\tau_p(z_1-z_2)(u_{\pm}^{-1} z_1z_2+\tau_p^2)}{2(v^2 z_1z_2 - \tau_p(z_2+vz_1) + \tau_p^2)(v^2 z_1z_2 - \tau_p(1+v)z_1 + \tau_p^2)}.
\end{equation}

\begin{rmk}
The PT-invariant models of ref. \cite{MP}, branch 2A, which is also linked to the Izergin-Korepin model \cite{IK}, 
are obtained as particular cases of this one. More precisely, setting 
\beq
p=\frac{2k^2}{k^4-1} \,,\quad t_p=\frac{2k^4}{(k^6+\eps_1)(k^2-\eps_1)} \,,\quad t_2 = \mp e^{-\frac{i\pi}{4}(1-\eps_1)} \frac{2k}{k^6+\eps_1} \,,
\eeq
one recovers the branch 2A Hamiltonians $H^{\pm}_{2A}(k,\eps_1,\eps_2)$ of ref. \cite{MP}. 
Note that each branch 2A Hamiltonian is related to the two models corresponding to the choices 
$u_{\pm}$ with $d = \eps_1 k^{-2\eps_2}$ or $d = \eps_1 k^{2\eps_2}$ through the following parametrization
\beq
v = \frac{d}{d^2-d+1} \,,\quad u_- = -(d^2-d+1)^2 \,,\quad u_+ = -\frac{(d^2-d+1)^2}{d^4} \,.
\eeq
\end{rmk}

\subsubsection{Generalization of the Bariev model (gB)\label{sec:sol25}} %25
 This models appears to be a multi-parametric interpolation of three known models, one of them being the Bariev model, see remarks \ref{rmk:bariev} and \ref{rmk:martins}.
 
The parameters which are left free are $p, q, t_1, t_2, t_p$. The remaining parameters entering the off-diagonal part of the Hamiltonian are given by
\beq
s_1=J\,\frac{Jt_1^2t_p^2-pqt_2^2}{t_1t_2^2}\,,\quad \displaystyle s_2=J^2\,\frac{Jt_1^2t_p^2-pqt_2^2}{t_2^3}\,,\quad \displaystyle s_3=-J^2\,\frac{pt_1}{t_2}\,,\quad \displaystyle t_3=-J\,\frac{qt_2}{t_1}\,,\quad \displaystyle s_p=J\,\frac{t_1^2t_p}{t_2^2},
\eeq
while on the diagonal we get:
\begin{align}
& Y=\frac{p^2t_1^2t_2 + Jpqt_1t_2^2 + J^2q^2t_2^3 - J^2t_1^3t_p^2}{t_1^2t_2t_p}\,,\quad \displaystyle X_{22}=\frac{p^2t_1^2t_2 + Jpqt_1t_2^2 + J^2q^2t_2^3}{t_1^2t_2t_p}\,,\quad \displaystyle X_{11}=J^2\,\frac{t_1t_p}{t_2},\\
& X_{12}=\frac{p^2t_1^2t_2 + Jpqt_1t_2^2 + J^2q^2t_2^3 + t_1^3t_p^2}{t_1^2t_2t_p}\,,\quad \displaystyle X_{21}=\frac{p^2t_1^2t_2 + Jpqt_1t_2^2 + J^2q^2t_2^3 + Jt_1^3t_p^2}{t_1^2t_2t_p}.
\end{align}
The S-matrix depends only on the reduced parameters $\tau_p$, $\theta$ and $\mu$:
\[
S(z_1,z_2) = -\frac{\Lambda(z_1,z_2)}{\Lambda(z_2,z_1)}
\]
where
\begin{align}
\Lambda(z_1,z_2) &= J\mu^4\tau_p^2 z_1^2z_2^2 - \mu^2\tau_p\theta z_1z_2(z_1+z_2) - J^2\mu^3\tau_p z_1z_2^2 + (\mu-\theta)(\mu-J^2\theta)z_1z_2 \nonumber \\
& + J^2\mu^3\tau_p^2 z_2^2 - \mu^2\tau_p(z_1+z_2) - J\mu\tau_p\theta z_2 + \mu^2\tau_p^2 
\end{align}
and the decay coefficient $N$ reads
\begin{equation}
N(z_1,z_2) = \frac{\tau_2\tau_p\mu^2(z_1-z_2)(1+\mu z_1z_2)}{2\Lambda(z_2,z_1)}.
\end{equation}

When $\displaystyle q=J\,\frac{t_1^2 t_p^2}{p t_2^2}$ (i.e. $\theta = J\mu^2\tau_p^2$), the $S$-matrix and the decay coefficient $N$ simplify and one gets 
\begin{equation}
S(z_1,z_2) = -\frac{J\mu^2\tau_p^2 z_1z_2 - J^2\mu\tau_p z_2 + 1}{J\mu^2\tau_p^2 z_1z_2 - J^2\mu\tau_p z_1 + 1}
\end{equation}
and
\begin{equation}
N(z_1,z_2) = \frac{\tau_2\tau_p(z_1-z_2)(1+\mu z_1z_2)}{2(z_1-\tau_p)(z_2-\tau_p)(J\mu^2\tau_p^2 z_1z_2 - J^2\mu\tau_p z_1 + 1)}\,.
\end{equation}

\begin{rmk}\label{rmk:bariev}
The Bariev model \cite{AB,AN} is obtained as a particular case of this one. More precisely, setting $p=q=1$, $J=\jmath$, $\displaystyle t_1=-\jmath^2\,\sqrt{t_p^2-1}$, $\displaystyle t_2=\jmath\sqrt{t_p^2-1}$, where $\jmath=\exp(\frac{2i\pi}{3})$, one gets for the other parameters $t_3=s_3=1$, $s_p=t_p$, $X_{11}=-t_p$, $\displaystyle Y=t_p+\frac{1}{t_p}$, $\displaystyle X_{12}=-\jmath t_p+\frac{1}{t_p}$, $\displaystyle X_{21}=-\jmath^2t_p+\frac{1}{t_p}$, $\displaystyle X_{22}=\frac{1}{t_p}$ which are the values of ref. \cite{AN}.
In that case, the S-matrix reads
\begin{equation}
S(z_1,z_2) = -\frac{\tau_p^2z_1^2z_2^2-\tau_pz_1^2z_2+z_1z_2-\tau_p^2z_2^2-\tau_pz_1+\tau_p^2}{\tau_p^2z_1^2z_2^2-\tau_pz_1z_2^2+z_1z_2-\tau_p^2z_1^2-\tau_pz_2+\tau_p^2}
\end{equation}
and the decay coefficient $N$ takes the form
\begin{equation}
N(z_1,z_2) = \frac{\tau_2\tau_p(z_1-z_2)(1-\jmath z_1z_2)}{2(\tau_p^2z_1^2z_2^2-\tau_pz_1z_2^2+z_1z_2-\tau_p^2z_1^2-\tau_pz_2+\tau_p^2)}.
\end{equation}
\end{rmk}

\begin{rmk}\label{rmk:martins}
The PT-invariant models of ref. \cite{MP}, branch 2B, are obtained as particular cases of this one. More precisely, setting 
\beq
p=q=-2i \,,\quad t_1=\pm \eps_1\eps_2e^{-2i\eps_2\pi/3} \,,\quad t_2=\pm \eps_1\eps_2e^{2i\eps_2\pi/3} \,,\quad t_p=-i\eps_1\sqrt{3}
\eeq
one recovers the branch 2B Hamiltonians $H^{\pm}_{2B}(\eps_1,\eps_2)$ of ref. \cite{MP} (with $\eps_1,\eps_2 \in \{-1,+1\}$). 

The main branch genus 5 model of ref. \cite{M13} is also a particular case of this one. More precisely, setting 
\beq
p=-\eps_2 \,,\quad q=-\eps_1 \,,\quad t_1=\frac{\sqrt{3} \mp i}{2}\,\sqrt{\eps_1\eps_2-1} \,,\quad t_2=\frac{\sqrt{3} \pm i}{2}\,\sqrt{\eps_1\eps_2-1}
\eeq
one recovers the main branch genus 5 Hamiltonians $H^{\pm}_{MB5}(\eps_1,\eps_2)$ of ref. \cite{M13} (we remind that here $\eps_1$ and $\eps_2$ are free parameters). 
\end{rmk}

\subsubsection{Generalization of the Hamiltonian built on $\cU_q(sl(2))$ special representation  at roots of unity (SpR)\label{sec:sol24}} %24

The parameters which are left free are $p, q, t_p, t_2, t_3$. The remaining parameters entering the off-diagonal part of the Hamiltonian are given by
\beq
 t_1=\frac{q t_2}{p}\,,\quad s_1=\frac{pt_3}{t_2}\,,\quad s_2=\frac{qt_3}{t_2}\,,\quad s_3=\frac{qt_3}{p}\,,\quad s_p=\frac{q(t_3^2 - t_3p+p^2)}{pt_p},\label{eq:sol24-1}
 \eeq
 while on the diagonal we get:
 \beq
X_{11}=0\,,\quad Y=X_{12}=X_{21}=X_{22}=\frac{t_3^2 - t_3p+p^2}{t_p}+\frac{q t_p}{p}.\label{eq:sol24-2}
\eeq

\begin{rmk}
Note that  eqs \eqref{eq:sol24-1} and \eqref{eq:sol24-2} look as if  some of the free parameters, say $t_p$,  cannot be set to zero. This is due to the choice of parametrization we made. However, one can choose
 alternative presentations. For instance, we could use as free parameters $p, q, s_p, t_1, s_3$. In that case, the remaining parameters take the form
\beq
 t_2=\frac{p t_1}{q}\,,\quad s_1=\frac{ps_3}{t_1}\,,\quad s_2=\frac{qs_3}{t_1}\,,\quad t_3=\frac{ps_3}{q}\,,\quad t_p=\frac{p(s_3^2 - s_3q+q^2)}{qs_p},\label{eq:sol24alt-1}
 \eeq
 and
 \beq
X_{11}=0\,,\quad Y=X_{12}=X_{21}=X_{22}=\frac{s_3^2 - s_3q+q^2}{s_p}+\frac{p s_p}{q}.\label{eq:sol24alt-2}
\eeq
 Clearly,  \eqref{eq:sol24alt-1} shows that we can now set $t_p=0$ without any problem. This new choice of parametrization is in fact the image under parity of the previous choice.
 \end{rmk}

The S-matrix depends only on the reduced parameters $\tau_p$ and $\tau_3$:
\begin{equation}
S(z_1,z_2) = -\frac{(\tau_3^2-\tau_3+1)z_1z_2 - \tau_p(z_1+z_2-\tau_3 z_2) + \tau_p^2}{(\tau_3^2-\tau_3+1)z_1z_2 - \tau_p(z_1+z_2-\tau_3 z_1) + \tau_p^2} 
\end{equation}
and the decay coefficient $N$ reads
\begin{equation}
N(z_1,z_2) = \frac{\tau_2\tau_p(z_1-z_2)}{2((\tau_3^2-\tau_3+1)z_1z_2 - \tau_p(z_1+z_2-\tau_3 z_1) + \tau_p^2)}.
\end{equation}

\begin{rmk}
The PT-invariant models of ref. \cite{MP}, branch 1B, which are linked to the models associated to 
special representation of $\cU_q(sl(2))$ at roots of unity \cite{Cou,GRS,GS}, 
are obtained as particular cases of this one. More precisely, setting 
\beq
p=q=-2i \,,\quad t_2=\pm 2 \,,\quad t_3=2i \,,\quad t_p=-2i\eps_1\sqrt{3}
\eeq
one recovers the branch 1B Hamiltonians $H^{\pm}_{1B}(\eps_1)$ of ref. \cite{MP}. 
\end{rmk}

\subsection{Generalization of the special branch genus 5 model (SB$_5$)\label{sec:sol14}} %14

The parameters which are left free are $p, q, t_2, Y$. The remaining parameters entering the off-diagonal part of the Hamiltonian are given by
\beq
t_p=s_p=0 \,,\quad t_1=\frac{q t_2}{p} \,,\quad s_1=-J^2\,\frac{p^2}{t_2} \,,\quad s_2=-J\,\frac{pq}{t_2} \,,\quad t_3=-J^2 p \,,\quad s_3=-Jq
\eeq
while on the diagonal we get:
\beq
X_{11}=0 \,,\quad X_{12}=X_{21}=X_{22}=Y.
\eeq
The S-matrix depends only on the reduced parameters $\theta$ and $\Upsilon$:
\begin{equation}
S(z_1,z_2) = -\frac{\theta z_1z_2(z_1-J^2z_2) - \Upsilon z_1z_2 + z_1-Jz_2}{\theta z_1z_2(z_2-J^2z_1) - \Upsilon z_1z_2 + z_2-Jz_1}
\end{equation}
and the decay coefficient $N$ reads
\begin{equation}
N(z_1,z_2) = -\frac{\tau_2(z_1-z_2)(\theta z_1z_2+1)}{2(\theta z_1z_2(z_2-J^2z_1) - \Upsilon z_1z_2 + z_2-Jz_1)}\,.
\end{equation}

\begin{rmk}
The special branch genus 5 model of ref. \cite{M13} is also a particular case of this one. More precisely, setting 
\beq
p=t_2=e^{\mp 2i\pi/3} \,,\quad q=-1 \,,\quad Y=4\Lambda
\eeq
one recovers the special branch genus 5 Hamiltonians $H^{\pm}_{SB5}(\Lambda)$ of ref. \cite{M13}. 
\end{rmk}

\subsection{Other models (17- and 14- vertex models)\label{sec:Ham02}}

The terminology ``17-vertex'' and ``14-vertex'' follows the one used for 19-vertex, see explanation in remark \ref{rmk:19v}.

As explained above, eqs. \eqref{E21}, \eqref{E12} and \eqref{E22} provide a set of constraints (denoted ${\cal C}_{|\Omega\rangle}$ hereafter) on the parameters entering the Hamiltonian. Solving these equations, we get a set of solutions, each of them determining an Hamiltonian solvable through CBA. Then,
 the BAE \eqref{eq:BAE} allows us to compute the eigenvalues \eqref{eq:energie} and eigenvectors \eqref{eq:psiM} of the model using the $S$ matrix and the decay coefficient $N$. The construction is based on the choice of a particular eigenvector of $H$: the pseudo-vacuum. 
 
Since we perform a classification, one shall get the same set of solutions whatever the choice of the pseudo-vacuum. In the case of the three-state Hamiltonian we are studying, there are two pseudo-vacua $|\Omega\rangle = \bigotimes_{i=1}^L |0\rangle$ and $|\widetilde\Omega\rangle = \bigotimes_{i=1}^L |2\rangle$.
Deploying the CBA machinery for each pseudo-vacuum leads to two distinct sets of constraint equations, ${\cal C}_{|\Omega\rangle}$ and ${\cal C}_{|\widetilde\Omega\rangle}$, the latter being obtained by applying the charge conjugation\footnote{In fact,  \eqref{eq:chargeconj} is just the gauge transformation relating $|\Omega\rangle$ to $|\widetilde\Omega\rangle$.} \eqref{eq:chargeconj} to the former. It follows in light of the foregoing that each solution of ${\cal C}_{|\Omega\rangle}$ should satisfy the equations coming from ${\cal C}_{|\widetilde\Omega\rangle}$. As it can be checked in Table \ref{table:actions}, it is indeed the case for the previous models.

However,  there are cases where a solution to  ${\cal C}_{|\Omega\rangle}$ does not solve identically ${\cal C}_{|\widetilde\Omega\rangle}$, but rather leads to additional constraints on the parameters. At this stage, the additional constraints could be interpreted as a failure in the CBA method: eigenvectors built on ${\cal C}_{|\widetilde\Omega\rangle}$ are not eigenvectors of the Hamiltonian based on a solution of ${\cal C}_{|\Omega\rangle}$. In fact, it just indicates that the eigenvectors obtained by CBA on ${\cal C}_{|\Omega\rangle}$ do not provide a complete basis of eigenvectors. One needs to consider a second pseudo-vacuum $|\widetilde\Omega\rangle$ to get a (tentatively) complete basis. Thus, it is the full sets of constraints ${\cal C}_{|\Omega\rangle}$ and ${\cal C}_{|\widetilde\Omega\rangle}$ that need to be considered. That is what we did for the class of models presented in this section. 
In practice, it is simpler, but equivalent, to apply the transformation \eqref{eq:chargeconj} to a given solution to the initial constraints ${\cal C}_{|\Omega\rangle}$, to impose the transformed solution to be also a solution of ${\cal C}_{|\Omega\rangle}$ (hence leading to more constraints on the parameters) and then to pull back the charge conjugation transformation \eqref{eq:chargeconj} on the result to get the final answer.

\subsubsection{Model $17V_1$\label{sec:sol17}}

Solving the constraints ${\cal C}_{|\Omega\rangle}$, the parameters which are left free are $p, q, t_p, t_2, t_3, s_3, X_{22}$ and the constraints on the parameters are given by
\beq
s_1=s_2=0\,,\quad s_p=\frac{pq}{t_p}\,,\quad t_1=\frac{q t_2}{p}
\eeq
and
\beq
X_{11}=0\,,\quad Y=\frac{p^2}{t_p}+\frac{q t_p}{p}\,,\quad X_{12}=\frac{p^2}{t_p}+\frac{q t_p}{p}+\frac{p t_3}{t_p}\,,\quad X_{21}=\frac{p^2}{t_p}+\frac{q t_p}{p}+\frac{t_p s_3}{p}\,.
\eeq
Solving now the constraints ${\cal C}_{|\widetilde\Omega\rangle}$, one gets two inequivalent possibilities:

\begin{itemize}
\item \textbf{Model $17V_{1a}$:} %17a
The additional conditions are
\beq
s_3=\epsilon q\,,\quad t_3=\epsilon p\,,\quad X_{22}=(1+\epsilon)Y,
\mb{with} \epsilon=\pm1,
\eeq
the diagonal terms reading now  
\beq
X_{12}=Y+\epsilon\,\frac{p^2}{t_p}\,,\quad X_{21}=Y+\epsilon\,\frac{q t_p}{p}.
\eeq

\item \textbf{Model $17V_{1b}$:} %17b
The additional conditions are
\beq
q = \frac{I p^3}{t_p^2}\,,\quad t_3=Ip \,,\quad s_3=\frac{p^3}{t_p^2}\,,\quad X_{22}=(1+I)\,\frac{p^2}{t_p}\,,
\eeq
 where $I$ is one solution of $I^2+1=0$. It leads to a redefinition of the parameters:
\beq
s_p= \frac{I p^4}{t_p^3}\,,\quad t_1=\frac{I p^2 t_2}{t_p^2} \,,\quad 
Y=(1+I)\,\frac{p^2}{t_p}\,,\quad X_{12}=(2I+1)\,\frac{p^2}{t_p}\,,\quad X_{21}=(I+2)\,\frac{p^2}{t_p}.
\eeq
\end{itemize}

For both models $17V_{1a}$ and $17V_{1b}$ the $S$-matrix is trivial, $S(z_1,z_2) = -1$, and the decay coefficient $N$ reads
\beq
N(z_1,z_2) = \frac{\tau_2\tau_p(z_1-z_2)}{2(z_1-\tau_p)(z_2-\tau_p)}.
\eeq

\subsubsection{Model $17V_2$\label{sec:sol21}} %21

Solving the constraints ${\cal C}_{|\Omega\rangle}$, the parameters which are left free are $p, q, t_p, t_2, t_3, s_3$ and the constraints on the parameters are given by
\beq
s_1=s_2=0\,,\quad s_p=\frac{pq}{t_p}\,,\quad t_1=-\frac{p^2 t_2}{t_p^2}
\eeq
and
\beq
X_{11}=Y=\frac{p^2}{t_p}+\frac{q t_p}{p}\,,\quad X_{12}=2Y-\frac{q t_p t_3}{p^2}\,,\quad X_{21}=2Y-\frac{p^2 s_3}{q t_p}\,,\quad X_{22}=2Y.
\eeq
Solving now the constraints ${\cal C}_{|\widetilde\Omega\rangle}$, the additional conditions are
\beq
s_3=q\,,\quad t_3=p\,,
\eeq
the diagonal terms reading now
 \beq
X_{12}=\frac{2p^2}{t_p}+\frac{q t_p}{p}\,,\quad X_{21}=\frac{p^2}{t_p}+\frac{2q t_p}{p}\,.
\eeq

The S-matrix depends only on the reduced parameters $\tau_p$ and $\theta$:
\begin{equation}
S(z_1,z_2) = -\frac{\theta\tau_p z_1z_2 - (\theta\tau_p^2+1)z_2+\tau_p}{\theta\tau_p z_1z_2 - (\theta\tau_p^2+1)z_1+\tau_p}
\end{equation}
and the decay coefficient $N$ reads
\begin{equation}
N(z_1,z_2) = \frac{-\tau_2(z_1-z_2)(z_1z_2-\tau_p^2)}{2(\theta\tau_p z_1z_2 - (\theta\tau_p^2+1)z_1+\tau_p)(z_1-\tau_p)(z_2-\tau_p)}.
\end{equation}

\subsubsection{Model $14V_1$\label{sec:sol3}} %3

Solving the constraints ${\cal C}_{|\Omega\rangle}$, the parameters which are left free are $p, t_p, t_2, t_3, X_{21}, X_{22}$ and the constraints on the parameters are given by
\beq
q=s_1=s_2=s_3=s_p=0\,,\quad t_1=-\displaystyle\frac{p^2 t_2}{t_p^2}
\,,
\quad
X_{11}=Y=\displaystyle\frac{p^2}{t_p}\,,\quad X_{12}=\displaystyle\frac{2p^2}{t_p}.
\eeq
Solving now the constraints ${\cal C}_{|\widetilde\Omega\rangle}$, the additional conditions are
\beq
t_3=\epsilon p\,,\quad X_{21}=X_{22}-\frac{p^2}{t_p}\mb{with} \epsilon=\pm1\,.
\eeq
The S-matrix and the decay coefficient $N$ read:
\begin{equation}
S(z_1,z_2) = -\frac{z_2-\tau_p}{z_1-\tau_p}
\,,
\quad
N(z_1,z_2) = \frac{\tau_2(z_1-z_2)(z_1z_2-\tau_p^2)}{2(z_1-\tau_p)^2(z_2-\tau_p)}.
\end{equation}

\subsubsection{Model $14V_2$\label{sec:sol4}} %4

Solving the constraints ${\cal C}_{|\Omega\rangle}$, the parameters which are left free are $p, t_p, t_1, t_2$ and the constraints on the parameters are given by
\beq
q=s_1=s_2=s_3=s_p=0\,,\quad t_3=-\frac{t_p^2 t_1}{p t_2}
\eeq
and
\beq
X_{11}=0\,,\quad X_{12}=Y=\frac{p^2}{t_p}\,,\quad X_{21}=X_{22}=\frac{p^2 t_2-t_p^2 t_1}{t_p t_2}.
\eeq
Solving now the constraints ${\cal C}_{|\widetilde\Omega\rangle}$, leads to one  additional condition
\beq
t_1=\frac{p^2 t_2}{t_p^2}
\eeq
which gives
$t_3=-p$ and $X_{21}=X_{22}=0$.

The S-matrix is trivial, $S(z_1,z_2) = -1$, and the decay coefficient $N$ reads:
\beq
N(z_1,z_2) = \frac{\tau_2(z_1-z_2)(z_1z_2+\tau_p^2)}{2\tau_p(z_1-\tau_p)(z_2-\tau_p)}.
\eeq

\section{Reduced Hamiltonians\label{sec:Hred}}
In this section, we use telescoping terms and gauge transformations (see section \ref{sec:sym}) to get a simple expression 
$\wt H$ for the Hamiltonians described in section \ref{sec:sol}.  
For all Hamiltonians, in a first step, we change the normalization and perform a gauge transformation: 
\beq
H_{red}= \cN_0\ G\otimes G\, \Big(H-\frac{V}{2}\,(s^z_1+s^z_2)\Big)\,G^{-1}\otimes G^{-1} \,,
\label{def:Hred}
\eeq
where $\cN_0$ is a constant, and $G$ a $3\times3$ diagonal matrix. Their exact form depends on the model we consider.
 We get in this way Hamiltonians that depend only on the physical parameters. The transformation is  valid 
only for generic values of the free parameters, see remark \ref{rmk:zero} below.

\subsection{Nineteen vertices}

\subsubsection{Generalized Zamolodchikov-Fateev model} %19

{From} the Hamiltonian $H$ given in section \ref{sec:sol19}, we perform the transformation \eqref{def:Hred} with 
\beq
\cN_0= \frac{t_p}{p^2}
\mb{and} G=\diag\Big(1,\left(\frac{t_2\,t_p^2}{s_1\,p^2}\right)^{1/4},1\Big).
\eeq
It leads to an Hamiltonian $H_{red}$ depending on $\tau_p$ and $\sigma$ only.

\begin{rmk}\label{rmk:zero}
Note that this transformation is not valid for $s_1=0$, although the ``raw'' Hamiltonian is well-defined in this case, see section \ref{sec:sol19}. This is due to the fact that the physical parameter $\sigma$ vanishes for this particular value of $s_1$. One can however work on the raw Hamiltonian to get then a reduced Hamiltonian containing only the physical parameter $\tau_p$.
\end{rmk}

To compare with existing models, we furthermore modify it to 
\beq
\wt H = \frac{-2 k^2}{k^4-1}\,H_{red} -  \frac{k^4+1}{k^4-1}\,(s^z_1+s^z_2)
\mb{with} \sigma=\left(\frac{k^2+1}{k}\right)^2
\eeq
we get 
\beq
\wt H=
 \left[ \begin {array}{ccccccccc} 0&0&0&0&0&0&0&0&0
\\ \noalign{\medskip}0&{\frac {{k}^{4}+1}{1-{k}^{4}}}&0&{\frac {2
\tau_p\,{k}^{2}}{1-{k}^{4}}}&0&0&0&0&0\\ \noalign{\medskip}0&0&{\frac 
{2\,{k}^{4}+2\,{k}^{2}+2}{1-{k}^{4}}}&0&{\frac {2k}{1-{k}^{2}}}&0&
{\frac {2{\tau_p}^{2}{k}^{2}}{1-{k}^{4}}}&0&0\\ \noalign{\medskip}0&
{\frac {2{k}^{2}}{ \left( 1-{k}^{4} \right) \tau_p}}&0&{\frac {{k}^{
4}+1}{1-{k}^{4}}}&0&0&0&0&0\\ \noalign{\medskip}0&0&{\frac {2k}{1-{k}
^{2}}}&0&{\frac {2\,{k}^{4}+2}{1-{k}^{4}}}&0&{\frac {2k{\tau_p}^{
2}}{1-{k}^{2}}}&0&0\\ \noalign{\medskip}0&0&0&0&0&{\frac {{k}^{4}+1}{1-
{k}^{4}}}&0&{\frac {2\tau_p\,{k}^{2}}{1-{k}^{4}}}&0
\\ \noalign{\medskip}0&0&{\frac {2{k}^{2}}{ \left(1- {k}^{4}
 \right) {\tau_p}^{2}}}&0&{\frac {2k}{{\tau_p}^{2} \left(1- {k}^{2}
 \right) }}&0&{\frac {2\,{k}^{4}+2\,{k}^{2}+2}{1-{k}^{4}}}&0&0
\\ \noalign{\medskip}0&0&0&0&0&{\frac {2{k}^{2}}{ \left(1- {k}^{4}
 \right) \tau_p}}&0&{\frac {{k}^{4}+1}{1-{k}^{4}}}&0
\\ \noalign{\medskip}0&0&0&0&0&0&0&0&0\end {array} \right] \,.
\label{htilde19}
\eeq
When $\tau_p=-1$, we  recover  the Zamolodchikov-Fateev model \cite{ZF} (or spin-1 XXZ spin chain). When $\tau_p=-\eps_1$, $k=\exp(\frac{\gamma}2+\frac{i\pi}{4}(1-\eps_1))$ and $\eps_1=\pm1$ is left free, we get
 the models ``branch 1A'' described in \cite{MP}. The models 7 and 10 of \cite{ITA} are also obtained in the same way. \\
To be complete, let us add that for $\tau_p=-1$, the Hamiltonian \eqref{htilde19} is related to the one based on $\cU_q(B^{(1)}_1)$ given in \cite{J86} by $H^{\cU_q(B_1^{(1)})}(1/k^2) = \wt H(k) + (\II\otimes e_{22}-e_{22}\otimes\II) + 2(\II\otimes e_{33}-e_{33}\otimes\II)$ (the $R$-matrix of $\cU_q(B_1^{(1)})$ we consider is normalized such that $R_{11}^{11}=1$).

\subsubsection{Generalized Izergin-Korepin model} %27

{From} the Hamiltonian $H$ given in section \ref{sec:sol27}, the transformation \eqref{def:Hred} with 
\beq
\cN_0= \frac{t_p}{p^2}
\mb{and} G=\diag\Big(1,\sqrt{\frac{t_2}{p}}\left(\frac{vu_-}{v-1}\right)^{1/4},1\Big)
\eeq
 leads to an Hamiltonian $H_{red}$ depending on $\tau_p$ and $v$ only. To compare with existing model, we first make a change of variable $\displaystyle v=\frac{k}{k^2-k+1}$, $u_-=-(k^2-k+1)^2$ and define
\begin{align}
\begin{split}
\wt H =&\  \frac{1}{(k^2-1)(k^2-k+1)}\,\Big( -(k^2-k+1)^2 H_{red} + \half (k^2+1)(k^2-k+1)\,(s^z_1+s^z_2)\\
&+\half (k^2-1)(k^2-k+1)\,(\II\otimes e_{22}-e_{22}\otimes\II) + \half (k-1)^3(k+1)\,(\II\otimes e_{33}-e_{33}\otimes\II) \Big)
\end{split}
\end{align}
that is, with $\tau'_p=\tau_p/v$,
\beq
\wt H= \left[ \begin {array}{ccccccccc} 
0 & 0 & 0 & 0 & 0 & 0 & 0 & 0 & 0
\\ 
\noalign{\medskip} 0 & \frac{k^2}{k^2-1} & 0 & \frac{-k\tau'_p}{k^2-1} & 0 & 0 & 0 & 0 & 0
\\ 
\noalign{\medskip} 0 & 0 & \frac{(k^3-k^2+1) k}{(k^3+1)(k-1)}  & 0 & \frac{-\tau'_p\sqrt{k}}{k^3+1} & 0 & \frac{-k^
2{\tau'_p}^2}{(k^3+1)(k-1)} & 0 & 0
\\ 
\noalign{\medskip} 0 & \frac{-k}{(k^2-1)\tau'_p} & 0 & \frac{1}{k^2-1} & 0 & 0 & 0 & 0 & 0
\\ 
\noalign{\medskip} 0 & 0 & \frac{-\sqrt{k}}{\tau'_p\,(k^3+1)} & 0 & \frac{k^3-k^2+k-1}{k^3+1} & 0 & \frac{k^{5/2}\tau'_p}{k^3+1
} & 0 & 0
\\ 
\noalign{\medskip} 0 & 0 & 0 & 0 & 0 & \frac{k^2}{k^2-1} & 0 & \frac{-k\tau'_p}{k^2-1} & 0 
\\ 
\noalign{\medskip} 0 & 0 & \frac{-k^2}{{\tau'_p}^2 (k^3+1)(k-1)} & 0 & \frac{k^{5/2}}{\tau'_p\, (k^3+1)} & 0 & \frac{k^3-k+1}{(k^3+1)(k-1)} & 0 & 0 
\\ 
\noalign{\medskip} 0 & 0 & 0 & 0 & 0 & \frac{-k}{(k^2-1)\tau'_p} & 0 & \frac{1}{k^2-1} & 0
\\
\noalign{\medskip} 0 & 0 & 0 & 0 & 0 & 0 & 0 & 0 & 0 
\end {array} \right] \,.
\eeq

For generic values of $k$ and $\tau_p$, the Hamiltonian $\wt H$ is  conjugated to the one based on $\cU_q(A_2^{(2)})$, and given in \cite{J86} (the $R$-matrix of $\cU_q(A_2^{(2)})$ we consider is normalized such that $R_{11}^{11}=1$):
\beq
\wt H(k)=\wt F\,H^{A_2^{(2)}}(k)\,\wt F^{-1}
\mb{with}
\wt F=\diag\big(u_1,\sqrt{u_1u_3},u_3\big) \otimes \diag\big(u_1\frac{v}{\tau_p},\sqrt{u_1u_3},u_3\frac{\tau_p}{v}\big).
\eeq
Note that the Hamiltonian $H^{A_2^{(2)}}(k)$ is related to the Izergin-Korepin model \cite{IK} through a constant gauge transformation and constant telescopic terms.

\medskip

To be complete, let us add that the Hamiltonian $H^{A_2^{(2)}}$ is related to the Branch 2A of \cite{MP} through the following transformation\footnote{We remind the correspondence with the notation of \cite{MP}: $u=exp(\lambda)$, $k=exp(\gamma/2+\frac{i\pi}{4}(1-\eps_1))$.}:
\beq
-2H^{A_2^{(2)}}(\eps_1 k^2) + (\II\otimes e_{22}-e_{22}\otimes\II) + 2\,(\II\otimes e_{33}-e_{33}\otimes\II) = F\,H^{2A}(k)\,F^{-1}
\label{branch2a}
\eeq
with
\beq
F=\diag\big(1,\frac{1}{\sqrt{\eps_1}},\sqrt{-\eps_2\eps_3},\sqrt{\eps_1},\frac{1}{\sqrt{-\eps_2\eps_3}},\sqrt{\eps_1},\sqrt{-\eps_2\eps_3},\frac{1}{\sqrt{\eps_1}},1\big).
\eeq
Note that in the correspondence \eqref{branch2a}, $H^{2A}$ corresponds to the branch 2A Hamiltonian $H_{12}$ of ref. \cite{MP} for $\eps_2=1$, while it corresponds to $H_{21}$ when $\eps_2=-1$.

\subsubsection{Generalized Bariev model} %25

{From} the Hamiltonian $H$ given in section \ref{sec:sol25}, we perform the transformation \eqref{def:Hred} with  
\beq
\cN_0 = \frac{1}{p}\sqrt{\frac{t_2}{t_1}}
\mb{and} G=\diag\Big(1,\sqrt{\frac{t_2}{p}},1\Big).
\eeq
We get an Hamiltonian $H_{red}$ depending on $\tau_p$, $\theta$ and $\mu$ only. 

This Hamiltonian can be related to the one of the Main Branch of ref. \cite{M13}, $H^{MB_5}$. One defines, with $\displaystyle\delta=\frac{\mu^2+J\theta\mu+J^2\theta^2}{4J^2\tau_p^2\mu^3}$,
\beq
\wt H = -\frac{J}{\tau_p\sqrt{\mu}}\,H_{red} + \delta\,(s^z_1+s^z_2-\II) + \half\,(J-J^{-1})\,(\II\otimes e_{33}-e_{33}\otimes\II)
\eeq
that is,
\beq
\wt H = \left[ \begin{array}{ccccccccc} 
-\delta &0&0&0&0&0&0&0&0 \\ 
\noalign{\medskip}0&0&0& -\frac{J}{\tau_p\mu} &0&0&0&0&0 \\ 
\noalign{\medskip}0&0& -\delta-J^2 &0& -\frac{J}{\tau_p\mu} &0& -J\mu^{-1} &0&0 \\ 
\noalign{\medskip}0& -\frac{J\theta}{\tau_p\mu} &0&0&0&0&0&0&0 \\ 
\noalign{\medskip}0&0& \frac{\theta-J\tau_p^2\mu^2}{\tau_p\mu} &0& \delta-1 &0& \frac{\theta-J\tau_p^2\mu^2}{J\tau_p\mu^2} &0&0 \\ 
\noalign{\medskip}0&0&0&0&0&0&0& \frac {J^2\theta}{\tau_p\mu^2} &0 \\ 
\noalign{\medskip}0&0& -J^2\mu &0& -\frac{J}{\tau_p} &0& -\delta-J &0&0 \\ 
\noalign{\medskip}0&0&0&0&0& \tau_p^{-1} &0&0&0 \\ 
\noalign{\medskip}0&0&0&0&0&0&0&0& -\delta
\end{array} \right] \,.
\eeq
Then one gets
\beq
\wt H = F\,H^{MB_5}\,F^{-1}  
\eeq
with $F=U\otimes U'$ where $U$ and $U'$ are expressed in terms of the free parameters $u_1,u_3$:
\beq
U=\diag\big(u_1,Z\left(\frac{\theta-J\tau_p^2\mu^2}{\mu}\right)^{1/4}\sqrt{u_1u_3},u_3)
\eeq
and
\beq
U'=\diag\big(-iJ\sqrt{\mu}\,u_1,\frac{1}{Z}\left(\frac{\theta-J\tau_p^2\mu^2}{\mu}\right)^{1/4}\sqrt{u_1u_3},-\frac{u_3}{iJ\sqrt{\mu}}\big).
\eeq
The parameters $(\eps_1,\eps_2)$ entering into the definition of $H^{MB_5}$ are given by one of these relations:
\begin{align}
\begin{split}
& \eps_1 = \frac{i\theta}{\tau_p\mu^{3/2}}\,,\quad \eps_2 = -\frac{iJ^2}{\tau_p\mu^{1/2}} \mb{with the choice} Z=1 \\
& \eps_1 = -\frac{iJ^2}{\tau_p\mu^{1/2}}\,,\quad \eps_2 = \frac{i\theta}{\tau_p\mu^{3/2}} \mb{with the choice} Z=\left(-\frac{J\theta}{\mu}\right)^{1/2}
\end{split}
\end{align}
and $J$ is related to the parameter $\gamma_0=\eps\frac{i\pi}{6}$ ($\eps=\pm 1$) of $H^{MB_5}$ by $J=-e^{-2\gamma_0}$.

\subsubsection{Generalization of the Hamiltonian associated to special representation of $\cU_q(sl(2))$ at roots of unity} %24

{From} the Hamiltonian $H$ given in section \ref{sec:sol24}, the transformation \eqref{def:Hred} with  
\beq
\cN_0= \frac{t_p}{p^2}
\mb{and} G=\diag\Big(1, \sqrt{\frac{t_2}{p}},1\Big)
\label{Hsymp24}
\eeq
produces an Hamiltonian $H_{red}$ depending on $\tau_p$, $\theta$ and $\tau_3$ only, that is, with $\delta=\tau_3^2-\tau_3+1+\tau_p^2\theta$,
\beq
H_{red} = \left[ \begin {array}{ccccccccc} 
0 & 0 & 0 & 0 & 0 & 0 & 0 & 0 & 0
\\ 
\noalign{\medskip} 0 & 0 & 0 & \tau_p & 0 & 0 & 0 & 0 & 0
\\ 
\noalign{\medskip} 0 & 0 & \frac{1}{2}\delta & 0 & \tau_p & 0 & \tau_p^2 & 0 & 0
\\ 
\noalign{\medskip} 0 & \tau_p\theta & 0 & 0 & 0 & 0 & 0 & 0 & 0
\\ 
\noalign{\medskip} 0 & 0 & \tau_p\tau_3\theta & 0 & 0 & 0 & \tau_p\tau_3 & 0 & 0
\\ 
\noalign{\medskip} 0 & 0 & 0 & 0 & 0 & \frac{1}{2}\delta & 0 & \tau_p\tau_3 & 0 
\\ 
\noalign{\medskip} 0 & 0 & \theta(\tau_3^2-\tau_3+1) & 0 & \tau_p\theta & 0 & \frac{1}{2}\delta & 0 & 0 
\\ 
\noalign{\medskip} 0 & 0 & 0 & 0 & 0 & \tau_p\tau_3\theta & 0 & \frac{1}{2}\delta & 0
\\
\noalign{\medskip} 0 & 0 & 0 & 0 & 0 & 0 & 0 & 0 & \delta
\end {array} \right] \,.
\eeq
The Branch 1B Hamiltonians of ref. \cite{MP} can be related to the Hamiltonian \eqref{Hsymp24} through the following transformation:
\beq
\frac{2(k^2-e^{2\gamma_0})}{(k^2-1)(1-e^{2\gamma_0})}\,H_{red} - \frac{k^2+1}{k^2-1}\,(s^z_1+s^z_2) = F\,H^{1B}(k,\eps_1,\eps_2)\,F^{-1} 
\label{twist24}
\eeq
with $\gamma_0=\eps_1\frac{i\pi}{3}$, $(\eps_1,\eps_2) \in \{-1,+1\}$, and $F=U\otimes U'$ where $U$ and $U'$ are expressed in terms of the free parameters $u_1,u_3$:
\beq
U=\diag\big(u_1,\sqrt{\frac{u_1u_3}{\Delta}},u_3) \mb{and} U'=\diag\big(u_1\sqrt{\theta},\sqrt{\frac{u_1u_3}{\Delta}},\frac{u_3}{\sqrt{\theta}}\big)\,,
\eeq
$\displaystyle \Delta = \frac{\eps_2}{e^{\gamma_0/2}\sqrt{\tau_3}}$, and the parameters $\theta$, $\tau_p$ and $\tau_3$ are linked to $k$ by the relations
\beq
\tau_3 = \frac{e^{\gamma_0}(k^2-1)}{k^2-e^{2\gamma_0}} 
\mb{and}
\tau_p \sqrt{\theta} = \frac{k(1-e^{2\gamma_0})}{k^2-e^{2\gamma_0}}.
\label{reltwist24}
\eeq
Note that the transformation \eqref{twist24} only holds when the parameters $\theta$, $\tau_p$ and $\tau_3$ are related to $k$ by \eqref{reltwist24}. Hence this model appears as a generalization of the Branch 1B models of \cite{MP}.

\subsection{Generalization of the special branch genus 5 model} %14

{From} the Hamiltonian $H$ given in section \ref{sec:sol14}, we perform the  transformation \eqref{def:Hred} with  
\beq
\cN_0 = \frac{1}{Y}
\mb{and} G=\diag\Big(1, \sqrt{\frac{t_2}{p}},1\Big)
\label{Hsymp13}
\eeq
that leads to an Hamiltonian $H_{red}$ depending on $\theta$ and $\Upsilon$ only. 

The genus 5 Special Branch Hamiltonian of ref. \cite{M13} can be related to the Hamiltonian \eqref{Hsymp13}. Indeed, one defines
\beq
\wt H = \frac{\Upsilon}{4J\sqrt{-\theta}} \,\big( 4\,H_{red} - (s^z_1+s^z_2) + \II \big)
\eeq
with $\gamma_0=\pm\frac{i\pi}{3}$, $J=e^{-2\gamma_0}$, that is
\beq
\wt H = \left[ 
\begin{array}{ccccccccc} 
\frac{\Upsilon}{4J\sqrt{-\theta}} & 0 & 0 & 0 & 0 & 0 & 0 & 0 & 0
\\ 
\noalign{\medskip} 0 & 0 & 0 & \frac{1}{J\sqrt{-\theta}} & 0 & 0 & 0 & 0 & 0
\\ 
\noalign{\medskip} 0 & 0 & \frac{\Upsilon}{4J\sqrt{-\theta}} & 0 & \frac{1}{J\sqrt{-\theta}} & 0 & 0 & 0 & 0
\\ 
\noalign{\medskip} 0 & -\frac{\sqrt{-\theta}}{J} & 0 & 0 & 0 & 0 & 0 & 0 & 0
\\ 
\noalign{\medskip} 0 & 0 & \sqrt{-\theta} & 0 & -\frac{\Upsilon}{4J\sqrt{-\theta}} & 0 & \frac{-J}{\sqrt{-\theta}} & 0 & 0
\\ 
\noalign{\medskip} 0 & 0 & 0 & 0 & 0 & 0 & 0 & \frac{-J}{\sqrt{-\theta}} & 0 
\\ 
\noalign{\medskip} 0 & 0 & 0 & 0 & -\frac{\sqrt{-\theta}}{J} & 0 & \frac{\Upsilon}{4J\sqrt{-\theta}} & 0 & 0
\\ 
\noalign{\medskip} 0 & 0 & 0 & 0 &  0 & \sqrt{-\theta} & 0 & 0 & 0
\\ 
\noalign{\medskip} 0 & 0 & 0 & 0 & 0 & 0 & 0 & 0 & \frac{\Upsilon}{4J\sqrt{-\theta}}
\end{array} 
\right] .
\eeq
Then one gets
\beq
\epsilon \wt H = F\,H^{SB5}(\Lambda,e^{\gamma_0})\,F^{-1}  
\eeq
with $F=U\otimes U'$ where $U$ and $U'$ are expressed in terms of the free parameters $u_1,u_3$:
\beq
U=\diag\big(u_1,\sqrt{u_1u_3},u_3) \mb{and} U'=\diag\big(\eps J^2\sqrt{-\theta}\,u_1,\sqrt{u_1u_3},\frac{\eps u_3}{J^2\sqrt{-\theta}}\big)\,,
\eeq
the parameter $\Lambda$ being linked to $\Upsilon$ and $\theta$ by $\displaystyle \Lambda = \frac{\eps\Upsilon}{4J\sqrt{-\theta}}$ (with $\eps=\pm 1$).

\subsection{Other models}

\subsubsection{Model $17V_{1a}$} %17a

{From} the Hamiltonian $H$ given in section \ref{sec:sol17} (case $1a$), we perform the transformation \eqref{def:Hred} 
\beq
\cN_0= \frac{t_p}{p^2}
\mb{and} G=\diag\Big(1, \sqrt{\frac{t_2}{p}},1\Big)
\eeq
that leads to an Hamiltonian $H_{red}$ depending on $\theta$ and $\tau_p$ only, that is,
\beq
H_{red} = \left[ 
\begin{array}{ccccccccc} 
0&0&0&0&0&0&0&0&0 \\ 
\noalign{\medskip}0&0&0& \tau_p &0&0&0&0&0 \\ 
\noalign{\medskip}0&0& \frac{1+\theta\tau_p^2}{2} &0& \tau_p &0& \tau_p^2 &0&0 \\  
\noalign{\medskip}0& \theta\tau_p &0&0&0&0&0&0&0 \\ 
\noalign{\medskip}0&0&0&0&0&0&0&0&0 \\ 
\noalign{\medskip}0&0&0&0&0& \frac{1+\theta\tau_p^2}{2}+\eps &0& \eps\tau_p &0 \\ 
\noalign{\medskip}0&0& \theta &0& \theta\tau_p &0& \frac{1+\theta\tau_p^2}{2} &0&0 \\ 
\noalign{\medskip}0&0&0&0&0& \eps\theta\tau_p &0& \frac{1+\theta\tau_p^2}{2}+\eps\theta\tau_p^2 &0 \\ 
\noalign{\medskip}0&0&0&0&0&0&0&0& (1+\eps)(1+\theta\tau_p^2)
\end{array} 
\right] \,.
\eeq

\subsubsection{Model $17V_{1b}$} %17b

{From} the Hamiltonian $H$ given in section \ref{sec:sol17} (case $1b$), we perform the transformation \eqref{def:Hred} with  
\beq
\cN_0= \frac{t_p}{p^2}
\mb{and} G=\diag\Big(1, \sqrt{\frac{t_2}{p}},1\Big)
\eeq
that leads to an Hamiltonian $H_{red}$ depending on $\tau_p$ only, that is
\beq
H_{red} = \left[ 
\begin{array}{ccccccccc} 
0&0&0&0&0&0&0&0&0 \\ 
\noalign{\medskip}0&0&0& \tau_p &0&0&0&0&0 \\ 
\noalign{\medskip}0&0& \half\,(1+I) &0& \tau_p &0& \tau_p^2 &0&0 \\ 
\noalign{\medskip}0& I\tau_p^{-1} &0&0&0&0&0&0&0 \\ 
\noalign{\medskip}0&0&0&0&0&0&0&0&0 \\ 
\noalign{\medskip}0&0&0&0&0& \half\,(1+3I) &0& I\tau_p &0 \\ 
\noalign{\medskip}0&0& I\tau_p^{-2} &0& I\tau_p^{-1} &0& \half\,(1+I) &0&0 \\ 
\noalign{\medskip}0&0&0&0&0& \tau_p^{-1} &0& \half\,(3+I) &0 \\ 
\noalign{\medskip}0&0&0&0&0&0&0&0& 1+I
\end{array} 
\right] \,.
\eeq

\subsubsection{Model $17V_2$} %21

{From} the Hamiltonian $H$ given in section \ref{sec:sol21}, we perform the  transformation \eqref{def:Hred} with  
\beq
\cN_0= \frac{t_p}{p^2}
\mb{and} G=\diag\Big(1, \sqrt{\frac{t_2}{p}},1\Big)
\eeq
that leads to an Hamiltonian $H_{red}$ depending on $\theta$ and $\tau_p$ only, that is
\beq
H_{red} = \left[ 
\begin{array}{ccccccccc} 
0&0&0&0&0&0&0&0&0 \\ 
\noalign{\medskip}0&0&0& \tau_p &0&0&0&0&0 \\ 
\noalign{\medskip}0&0& \frac{1+\theta\tau_p^2}{2} &0& \tau_p &0& \tau_p^2 &0&0 \\ 
\noalign{\medskip}0& \tau_p\theta &0&0&0&0&0&0&0 \\ 
\noalign{\medskip}0&0&0&0& 1+\theta\tau_p^2 &0&0&0&0 \\ 
\noalign{\medskip}0&0&0&0&0& \frac{3+\theta\tau_p^2}{2} &0& \tau_p &0 \\ 
\noalign{\medskip}0&0& \theta &0& -\tau_p^{-1} &0& \frac{1+\theta\tau_p^2}{2} &0&0 \\ 
\noalign{\medskip}0&0&0&0&0& \tau_p\theta &0& \frac{1+3\theta\tau_p^2}{2} &0 \\ 
\noalign{\medskip}0&0&0&0&0&0&0&0& 2(1+\theta\tau_p^2)
\end{array} 
\right] \,.
\eeq

\subsubsection{Model $14V_1$} %3

{From} the Hamiltonian $H$ given in section \ref{sec:sol3}, we perform the  transformation \eqref{def:Hred} with  
\beq
\cN_0= \frac{t_p}{p^2}
\mb{and} G=\diag\Big(1, \sqrt{\frac{t_2}{p}},1\Big)
\eeq
that leads to an Hamiltonian $H_{red}$ depending on $\tau_p$, $\xi=X_{22}/p$ and $\eps=\pm 1$ only, that is
\beq
H_{red} = \left[ 
\begin{array}{ccccccccc} 
0&0&0&0&0&0&0&0&0 \\ 
\noalign{\medskip}0&0&0& \tau_p &0&0&0&0&0 \\ 
\noalign{\medskip}0&0& \half &0& \tau_p &0& \tau_p^2 &0&0 \\ 
\noalign{\medskip}0&0&0&0&0&0&0&0&0 \\ 
\noalign{\medskip}0&0&0&0& 1 &0&0&0&0 \\ 
\noalign{\medskip}0&0&0&0&0& \frac{3}{2} &0& \eps\tau_p &0 \\ 
\noalign{\medskip}0&0&0&0& -\tau_p^{-1} &0& \frac{1}{2} &0&0 \\ 
\noalign{\medskip}0&0&0&0&0&0&0& \tau_p\,\xi-\frac{3}{2} &0 \\ 
\noalign{\medskip}0&0&0&0&0&0&0&0& \tau_p\,\xi
\end{array} 
\right] \,.
\eeq

\subsubsection{Model $14V_2$} %4

{From} the Hamiltonian $H$ given in section \ref{sec:sol4}, we perform the transformation \eqref{def:Hred} with  
\beq
\cN_0= \frac{t_p}{p^2} 
\mb{and} G=\diag\Big(1, \sqrt{\frac{t_2}{p}},1\Big)
\eeq
that leads to an Hamiltonian $H_{red}$ depending on $\tau_p$ only, that is
\beq
H_{red} = \left[ 
\begin{array}{ccccccccc} 
0&0&0&0&0&0&0&0&0 \\ 
\noalign{\medskip}0&0&0& \tau_p &0&0&0&0&0 \\ 
\noalign{\medskip}0&0& \half &0& \tau_p &0& \tau_p^2 &0&0 \\ 
\noalign{\medskip}0&0&0&0&0&0&0&0&0 \\ 
\noalign{\medskip}0&0&0&0&0&0&0&0&0 \\ 
\noalign{\medskip}0&0&0&0&0& \half &0&- \tau_p &0 \\ 
\noalign{\medskip}0&0&0&0& \tau_p^{-1}&0& \half &0&0 \\ 
\noalign{\medskip}0&0&0&0&0&0&0& -\half &0 \\ 
\noalign{\medskip}0&0&0&0&0&0&0&0&0
\end{array} 
\right] \,.
\eeq

\section{Conclusion}
In this paper we have provided a classification of `all' the Hamiltonians with rank 1 symmetry and nearest neighbour interactions, 
acting on a periodic three-state spin chain, and solvable through (a generalization of) the coordinate Bethe ansatz (CBA). 

Of course, the search for an $R$-matrix formulation of the new models presented here should be done, but 
many directions of generalizations can also be planed. First of all, the case with rank 2 symmetry algebra can also be easily done using the same method. 
Next, the integrable Hamiltonians that are not solvable through CBA, such as the ones obtained from Temperley-Lieb algebras, should be classified too. 
Finally, a similar classification for models solvable through algebraic Bethe ansatz would help to give a better understanding of the connection between 
these two approaches.

There is also a natural question that arises from this classification: what possible extensions of this work can be envisioned for $n$-state Hamiltonians? 
A priori, the method becomes rather intricate when increasing the number of states on each site, so that there is few hope that this can be done in the same way. However, increasing the rank of the symmetry algebra at the same time could provide some simplification.
This question is of relevance to recent developments in ultracold gases in optical lattices, such as the achievement of cooling down to quantum degeneracy five Ytterbium isotopes\cite{Fuk} which exhibit an enlarged $SU(6)$ symmetry.

\section*{Acknowledgements}
We are grateful to Dominique Caron and Phares Chakour for technical support when constructing the web page \cite{web}.
We thank G.A.P. Ribeiro for pointing out the CBA for models based on Temperley-Lieb algebras.
%and the references \cite{CBA-TL}, \cite{TL-open} and \cite{TL}. 

\newpage 

\appendix
\section{ $P,C,T$ transformations\label{sec:table}}

\begin{table}[htbp]
\begin{center}
\begin{tabular}{|c|c|c|c|c|c|}
\hline
Model & \# vertices & P action & C action & T action & Invariances \\[.24ex]
\hline
gZF & 19 & gZF & gZF & gZF & $P$, $C$, $T$ \\[1ex]
gIK & 19 & gIK$\big|_{u_\pm \to u_\mp}$ & gIK$\big|_{u_\pm \to u_\mp}$ & gIK & $PC$, $T$ \\[1ex]
gB & 19 & gB$\big|_{J \to J^2}$ & gB$\big|_{J \to J^2}$ & gB & $PC$, $T$ \\[1.2ex]
SpR & 19 & SpR & SpR & SpR & $P$, $C$, $T$ \\[1ex]
\hline
SB$_5$ & 17 & SB$_5\big|_{J \to J^2}$ & $C(SB_5) = T(SB_5)\big|_{J \to J^2}$ & $T(SB_5)$ & $PCT$ \\[1ex]
$17V_{1a}$  & 17 & $17V_{1a}$ & $17V_{1a}$ & $T(17V_{1a})$ & $P$, $C$ \\[1ex]
$17V_{1b}$ & 17 & $17V_{1b}\big|_{I \to -I}$ & $C(17V_{1b})$ & $T(17V_{1b})$ & $-$ \\[1ex]
$17V_{2}$ & 17 & $17V_{2}$ & $17V_{2}$ & $T(17V_2)$ & $P$, $C$ \\[1ex]
\hline
$14V_1$ & 14 & $P(14V_1)$ & $C(14V_1) = P(14V_1)$ & $T(14V_1)$ & $PC$ \\[1ex]
$14V_2$ & 14 & $P(14V_2)$ & $C(14V_2) = P(14V_2)$ & $T(14V_2)$ & $PC$ \\[1ex]
\hline
\end{tabular}
\caption{Actions of $P,C,T$ on the Hamiltonians\label{table:PCT}}
\end{center}
\label{table:actions}
\end{table}%

\end{document}